\documentclass[aps,twocolumn,pra,superscriptaddress,showpacs,tightenlines]{revtex4-1}
\usepackage{multirow}
\usepackage{array}
\usepackage{amsmath}
\usepackage{graphicx}
\usepackage{color}
\usepackage{amsfonts}
\usepackage{txfonts}
\usepackage[colorlinks,citecolor=blue]{hyperref}
\begin{document}
\title{Multimode optomechanical cooling via general dark-mode control}
\author{Jian Huang}
\affiliation{Key Laboratory of Low-Dimensional Quantum Structures and Quantum Control of Ministry of Education, Key Laboratory for Matter Microstructure and Function of Hunan Province, Department of Physics and Synergetic Innovation Center for Quantum Effects and Applications, Hunan Normal University, Changsha 410081, China}
\author{Deng-Gao Lai}
\affiliation{Theoretical Quantum Physics Laboratory, RIKEN Cluster for Pioneering Research, Wako-shi, Saitama 351-0198, Japan}
\author{Cheng Liu}
\affiliation{Key Laboratory of Low-Dimensional Quantum Structures and Quantum Control of Ministry of Education, Key Laboratory for Matter Microstructure and Function of Hunan Province, Department of Physics and Synergetic Innovation Center for Quantum Effects and Applications, Hunan Normal University, Changsha 410081, China}
\author{Jin-Feng Huang}
\affiliation{Key Laboratory of Low-Dimensional Quantum Structures and Quantum Control of Ministry of Education, Key Laboratory for Matter Microstructure and Function of Hunan Province, Department of Physics and Synergetic Innovation Center for Quantum Effects and Applications, Hunan Normal University, Changsha 410081, China}
\author{Franco Nori}
\affiliation{Theoretical Quantum Physics Laboratory, RIKEN Cluster for Pioneering Research, Wako-shi, Saitama 351-0198, Japan}
\affiliation{RIKEN Center for Quantum Computing (RQC), 2-1 Hirosawa, Wako-shi, Saitama 351-0198, Japan}
\affiliation{Physics Department, University of Michigan, Ann Arbor, Michigan 48109-1040, USA}
\author{Jie-Qiao Liao}
\email{Corresponding author: jqliao@hunnu.edu.cn}
\affiliation{Key Laboratory of Low-Dimensional Quantum Structures and Quantum Control of Ministry of Education, Key Laboratory for Matter Microstructure and Function of Hunan Province, Department of Physics and Synergetic Innovation Center for Quantum Effects and Applications, Hunan Normal University, Changsha 410081, China}

\begin{abstract}
The dark-mode effect is a stubborn obstacle for ground-state cooling of multiple degenerate mechanical modes optomechanically coupled to a common cavity-field mode. Here we propose an auxiliary-cavity-mode method for simultaneous ground-state cooling of two degenerate or near-degenerate mechanical modes by breaking the dark mode. We find that the introduction of the auxiliary cavity mode not only breaks the dark-mode effect, but also provides a new cooling channel to extract the thermal excitations stored in the dark mode. Moreover, we study the general physical-coupling configurations for breaking the dark mode in a generalized network-coupled four-mode optomechanical system consisting of two cavity modes and two mechanical modes. We find the analytical dark-mode-breaking condition in this system. This method is general and it can be generalized to break the dark-mode effect and to realize the simultaneous ground-state cooling in a multiple-mechanical-mode optomechanical system. We also demonstrate the physical mechanism behind the dark-mode breaking by studying the breaking of dark-state effect in the $N$-type four-level atomic system. Our results not only provide a general method to control various dark-mode and dark-state effects in physics, but also present an opportunity to the study of macroscopic quantum phenomena and applications in multiple-mechanical-resonator systems.
\end{abstract}

\maketitle

\section{Introduction\label{introduce}}
Considerable recent interest in cavity optomechanics~\cite{Kippenberg2008,Aspelmeyer2014,Metcalfe2014} has been paid to multimode optomechanical systems involving two~\cite{Mancini2002,Borkje2011,Stannigel2012,Massel2012,Spethmann2016,Piergentili2018,Yang2020,Riedinger2018,Ockeloen-Korppi2018,Kotler2021,Lepinay2021} or multiple~\cite{Bhattacharya2008,Xuereb2012,Xu2013,Nielsen2017,Mari2013,Zhang2015,Mercade2021,Lai2020pra,Lai2021pra,Heinrich2011,Ludwig2013,Xuereb2014,Cernotik2018} mechanical resonators; in particular, the two-mechanical-mode optomechanical systems have been realized in several experimental platforms~\cite{Massel2012,Spethmann2016,Riedinger2018,Ockeloen-Korppi2018,Piergentili2018,Yang2020,Kotler2021,Lepinay2021}. The study of multiple-mechanical-mode optomechanical systems has significance in both fundamental quantum physics~\cite{Schwab2005} and modern quantum technologies~\cite{Metcalfe2014}. For example, generation of macroscopic mechanical entanglement in multimode optomechanical systems has been experimentally demonstrated~\cite{Riedinger2018,Ockeloen-Korppi2018,Kotler2021,Lepinay2021}. Multimode optomechanical systems have also been considered to study quantum many-body effects~\cite{Heinrich2011,Ludwig2013,Xuereb2014}, high performance sensors~\cite{Massel2011,Huang2013}, precise measurement~\cite{Peano2015}, and nonreciprocal phonon or photon transport~\cite{Xu2015,Xu2016,Shen2016,Fang2017,Malz2018,Shen2018,Mathew2018,Xu2019}.

The simultaneous ground-state cooling of multiple mechanical modes has become a desired task because it is a prerequisite for the manipulation of macroscopic mechanical coherence~\cite{Schwab2005}. In particular, owing to the inherent structural features in multiple mechanical-mode optomechanical systems, people prefer to implement the simultaneous cooling of multiple mechanical resonators, rather than to cool these resonators one by one using the single-resonator cooling techniques. Though great success has been made in cooling a single mechanical resonator in optomechanical systems~\cite{Wilson-Rae2007,Marquardt2007,Genes2008b,Chan2011,Teufel2011,Liu2013a,Xu2017,Clarkl2017,Qiu2020,Rossi2018,Tebbenjohanns2019,Guo2019}, it remains a great challenge to perform ground-state cooling of multiple degenerate mechanical resonators coupled to a common cavity field, due to the mechanical dark-mode effect~\cite{Genes2008a,Massel2012,Shkarin2014,Kuzyk2017,Ockeloen-Korppi2019}. For the two-mechanical-mode case, the dark-mode effect has been theoretically found~\cite{Genes2008a} and experimentally demonstrated~\cite{Ockeloen-Korppi2019}. Meanwhile, the dark mode formed in optomechanical systems involving two cavity modes and one mechanical mode has also been found~\cite{Dong2012,Wang2012,Tian2012,Lake2020}. So far, theoretical proposals for optomechanical cooling of multiple mechanical resonators coupled in-series have been proposed~\cite{Lai2018,Lai2021}, and cooling of multimodes in a resonator have been analyzed with the cold-damping feedback method~\cite{Sommer2019,Sommer2020}. In addition, ground-state cooling of multiple mechanical resonators has been proposed based on synthetic magnetism~\cite{Lai2020} and reservoir engineering~\cite{Naseem2021}. In particular, we mention that considerable attention~\cite{Miao2009,Elste2009,Xuereb2015,Burgwal2020} has been paid to the suppression of optomechanical backaction and to the improvement of restrictions for effective cooling in multimode optomechanical systems.

In this paper, we propose an auxiliary-cavity-mode method for breaking the dark mode and further realizing ground-state cooling of two degenerate mechanical modes. We also explore the general coupling configurations for breaking the dark mode in a generalized four-mode optomechanical system, in which all the two-node couplings exist among two cavity modes and two mechanical modes. We find the analytical general conditions for forming and breaking the dark mode. Moreover, we extend this method to break the dark modes and realize the ground-state cooling of the multiple mechanical resonators. Correspondingly, our scheme can also be used to cool multiple degenerate or near-degenerate vibrational modes in a resonator. We also describe the physical mechanism for breaking the dark-state effect in the $N$-type four-level atomic system. The physical mechanism of our scheme is general and it can be generalized to control other dark-mode and dark-state effects in various branches of physics. In this sense, our results not only open up a different route to the realization of simultaneous ground-state cooling of two degenerate or near-degenerate mechanical resonators and enrich the technique of few-body resonator cooling, but also initiate advances in dark-state engineering~\cite{Scullybook}.

The rest of this paper is organized as follows. In Sec.~\ref{model}, we introduce the physical model and present the quantum Langevin equations. In Sec.~\ref{cooling}, we study ground-state cooling of the two mechanical modes by calculating the final mean phonon numbers. In Sec.~\ref{Universal}, we obtain the universal dark-mode-breaking conditions in a four-mode optomechanical system consisting of two cavity modes and two mechanical modes. We also analyze the quantum interference effect in the energy-level transitions of the system. In Sec.~\ref{Coolmulti}, we study the simultaneous ground-state cooling of multiple mechanical modes and the dark-mode breaking in a multiple-mechanical-mode optomechanical system. In Sec.~\ref{mechanism}, we describe the physical mechanism for breaking the dark-state effect in the $N$-type four-level atomic system. We present a discussion of the experimental implementation of our scheme in Sec.~\ref{Discuss} and summarize this work in Sec.~\ref{Conclu}.

\section{Physical model and equations of motion \label{model}}

%%%%%%%%%%%%%%%%%%%%%%%%%%%%%%
\begin{figure}[tbp]
\center
\includegraphics[width=0.47 \textwidth]{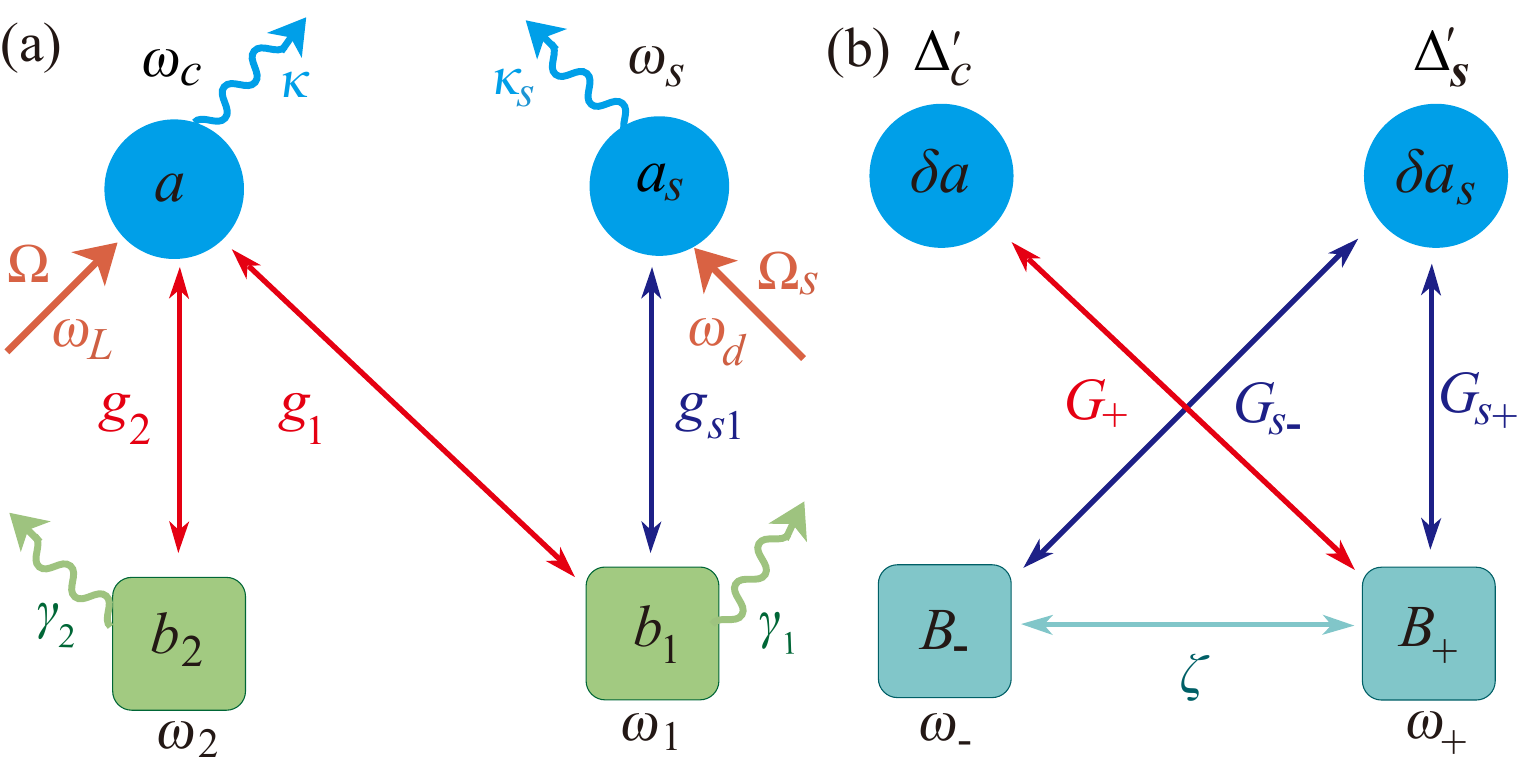}
\caption{(a) Schematic of the $N$-type four-mode optomechanical system. An intermediate coupling cavity mode $a$ with the resonance frequency $\omega_{c}$ is optomechanically coupled to two mechanical modes $b_{1}$ and $b_{2}$, with the corresponding resonance frequencies $\omega_{1}$ and $\omega_{2}$. An auxiliary cavity mode $a_{s}$ with the resonance frequency $\omega_{s}$ is optomechanically coupled to the mechanical mode $b_{1}$. The coupling strength between the cavity mode $a$ ($a_{s}$) and the mechanical mode $b_{l=1,2}$ ($b_{1}$) is denoted by $g_{l=1,2}$ ($g_{s1}$). The cavity mode $a$ ($a_{s}$) is driven with the driving frequency $\omega_{L}$ ($\omega_{d}$) and the driving amplitude $\Omega$ ($\Omega_{s}$). The decay rates of the intermediate cavity mode, the auxiliary cavity mode, and the two mechanical modes are denoted by $\kappa$, $\kappa_{s}$, and $\gamma_{l=1,2}$, respectively. (b) Coupling configuration associated with the approximate linearized Hamiltonian~(\ref{Hamitthreedig}). The auxiliary cavity mode $\delta a_{s}$ (with driving detuning $\Delta_{s}^{\prime}$) is coupled to the two hybrid mechanical modes $B_{\pm}$ (with resonance frequency $\omega_{\pm}$) via the effective coupling strengths $G_{s\pm}$. The cavity mode $\delta a$ (with driving detuning $\Delta_{c}^{\prime}$) is coupled to the hybrid mechanical mode $B_{+}$  via the effective coupling strength $G_{+}$. The phonon-hopping interaction between the two hybrid mechanical modes is denoted by $\zeta$. In the degenerate-mechanical-mode case and in the absence of auxiliary cavity, i.e., $\zeta=0$ and $G_{s\pm}=0$, the hybrid mechanical mode $B_{-}$ is decoupled from both the cavity mode $\delta a$ and the hybrid mechanical mode $B_{+}$.}
\label{Fig1}
\end{figure}
%%%%%%%%%%%%%%%%%%%%%%%%%%%%%%

We consider an $N$-type four-mode optomechanical system consisting of two cavity modes (an intermediate cavity mode $a$ and an auxiliary cavity mode $a_{s}$) and two mechanical modes ($b_{1}$ and $b_{2}$), as shown in Fig.~\ref{Fig1}(a). Here, the intermediate cavity mode is coupled to the two mechanical modes via radiation-pressure interactions. When the frequencies of the two mechanical modes are degenerate, a dark mode is formed in this linearized optomechanical system~\cite{Genes2008b}. This dark mode decouples from the intermediate cavity mode, and hence the ground-state cooling of the two mechanical modes is largely suppressed. To break the dark-mode effect, we introduce the auxiliary cavity mode $a_{s}$, which is optomechanically coupled to the mechanical mode $b_{1}$. Moreover, two driving fields are applied to the cavity modes to control the optical and mechanical degrees of freedom. In a rotating frame defined by the operator $\exp[-i(\omega_{L}a^{\dagger}a+\omega_{d}a_{s}^{\dagger}a_{s})t]$, the system Hamiltonian is given by ($\hbar=1$)
\begin{eqnarray}
H_{I}&=&\Delta_{c}a^{\dagger}a+\Delta_{s}a_{s}^{\dagger}a_{s}+\sum_{l=1,2}[\omega_{l}b_{l}^{\dagger}b_{l}+g_{l}a^{\dagger}a(b_{l}^{\dagger}+b_{l})]\nonumber\\
&&+g_{s1}a_{s}^{\dagger}a_{s}(b_{1}^{\dagger}+b_{1})+(\Omega a^{\dagger}+\Omega _{s}a_{s}^{\dagger }+\mathrm{H.c.}), \label{Hamit1}
\end{eqnarray}
where $\Delta_{c}=\omega_{c}-\omega_{L}$ ($\Delta_{s}=\omega_{s}-\omega_{d}$) is the driving detuning of the cavity frequency $\omega_{c}$ ($\omega_{s}$) with respect to its driving frequency $\omega_{L}$ ($\omega_{d}$); $a$ $(a^{\dagger})$, $a_{s}$ $(a_{s}^{\dagger})$, and $b_{l=1,2}$ $(b^{\dagger}_{l})$ are, respectively, the annihilation (creation) operators of the intermediate cavity mode, the auxiliary cavity mode, and the $l$th mechanical mode, with the corresponding resonance frequencies $\omega_{c}$, $\omega_{s}$, and $\omega_{l}$. The $g_{l=1,2}$ $(g_{s1})$ term describes the optomechanical coupling between the mechanical mode $b_{l=1,2}$ ($b_{1}$) and the cavity mode $a$ ($a_{s}$). The parameters $\omega_{L}$ ($\omega_{d}$) and $\Omega$ ($\Omega_{s}$) are the driving frequency and amplitude related to the driving field of the cavity mode $a$ ($a_{s}$), respectively.

To include the dissipations in this system, we assume that the two cavity fields are coupled to individual vacuum baths, and that the two mechanical modes are coupled to individual heat baths. Then the evolution of this system is governed by the quantum Langevin equations
\begin{subequations}
\begin{align}
\dot{a}=&-(i\Delta_{c}+\kappa)a-i\Omega-i\sum_{l=1,2}g_{l}a(b_{l}^{\dagger}+b_{l})+\sqrt{2\kappa}a_{\textrm{in}},\\
\dot{a}_{s}=&-(i\Delta_{s}+\kappa_{s})a_{s}-i[\Omega_{s}+g_{s1}a_{s}(b_{1}^{\dagger}+b_{1})]+\sqrt{2\kappa_{s}}a_{s,\textrm{in}},\\
\dot{b}_{1}=&-(\gamma_{1}+i\omega_{1})b_{1}-i(g_{1}a^{\dagger}a+g_{s1}a_{s}^{\dagger}a_{s})+\sqrt{2\gamma_{1}}b_{1,\textrm{in}}, \\
\dot{b}_{2}=&-(\gamma_{2}+i\omega_{2})b_{2}-ig_{2}a^{\dagger}a+\sqrt{2\gamma _{2}}b_{2,\textrm{in}},\\ \notag
\end{align}
\end{subequations}
where $\kappa$, $\kappa_{s}$, and $\gamma_{l=1,2}$ are the decay rates of the intermediate cavity mode $a$, the auxiliary cavity mode $a_{s}$, and the $l$th mechanical mode, respectively. The operators $a_{\textrm{in}}$ $(a^{\dagger}_{\textrm{in}})$, $a_{s,\textrm{in}}$ $(a^{\dagger}_{s,\textrm{in}})$, and $b_{l,\textrm{in}}$ $(b^{\dagger}_{l,\textrm{in}})$ are the noise operators associated with the intermediate cavity mode, the auxiliary cavity mode, and the $l$th mechanical mode, respectively. These noise operators have zero mean values and obey the correlation functions~\cite{Gardiner2000,Agarwal2013},
\begin{subequations}
\begin{align}
\langle a_{\textrm{in}}(t) a_{\textrm{in}}^{\dagger}(t^{\prime})\rangle&=\delta(t-t^{\prime}), \\
\langle a_{\textrm{in}}^{\dagger}(t) a_{\textrm{in}}(t^{\prime})\rangle&=0, \\
\langle a_{s,\textrm{in}}(t) a_{s,\textrm{in}}^{\dagger}(t^{\prime})\rangle&=\delta(t-t^{\prime}), \\
\langle a_{s,\textrm{in}}^{\dagger}(t) a_{s,\textrm{in}}(t^{\prime})\rangle&=0, \\
\langle b_{l,\textrm{in}}(t) b_{l,\textrm{in}}^{\dagger}(t^{\prime})\rangle&=(\bar{n}_{l}+1)\delta(t-t^{\prime}), \\
\langle b_{l,\textrm{in}}^{\dagger}(t) b_{l,\textrm{in}}(t^{\prime})\rangle&=\bar{n}_{l}\delta(t-t^{\prime}) \\ \notag
\end{align}
\end{subequations}
for $l=1,2$, where $\bar{n}_{l=1,2}$ is the average thermal-phonon occupation number associated with the $l$th mechanical mode.

To cool the mechanical modes, we assume that the two cavity modes are driven strongly and then the mean photon numbers in the two cavities are large enough and this four-mode optomechanical system can be processed by the linearization procedure. In this way, the operators $o\in\{a, a^{\dagger}, a_{s}, a_{s}^{\dagger}, b_{l}, b^{\dagger}_{l}\}$ can be expressed as a summation of steady-state average values and quantum fluctuation operators, i.e., $o=\langle o\rangle_{\text{ss}}+\delta o$. By separating the steady-state average values and the quantum fluctuation operators, we can obtain the linearized Langevin equations for the quantum fluctuation operators
\begin{subequations}
\label{linzedlang}
\begin{align}
\delta\dot{a}=&-(\kappa+i\Delta_{c}^{\prime})\delta a-i\sum_{l=1,2}[G_{l}(\delta b_{l}+\delta b_{l}^{\dagger})]+\sqrt{2\kappa}a_{\textrm{in}},  \\
\delta\dot{a}_{s} =&-(\kappa_{s}+i\Delta_{s}^{\prime})\delta a_{s}-iG_{s1}(\delta b_{1}+\delta b_{1}^{\dagger})+\sqrt{2\kappa _{s}}a_{s,\textrm{in}}, \\
\delta\dot{b}_{1}=&-(i\omega_{1}+\gamma_{1})\delta b_{1}-iG_{1}^{\ast}\delta a-iG_{1}\delta a^{\dagger}-iG_{s1}^{\ast}\delta a_{s}\nonumber \\
&-iG_{s1}\delta a_{s}^{\dagger}+\sqrt{2\gamma_{1}}b_{1,\textrm{in}}, \\
\delta\dot{b}_{2}=&-(i\omega_{2}+\gamma_{2})\delta b_{2}-iG_{2}^{\ast}\delta a-iG_{2}\delta a^{\dagger}+\sqrt{2\gamma_{2}}b_{2,\textrm{in}},
\end{align}
\end{subequations}
where the parameters $\Delta_{c}^{\prime}=\Delta_{c}+2g_{1}\text{Re}(\beta_{1}) +2g_{2}\text{Re}(\beta_{2})$ and $\Delta_{s}^{\prime}=\Delta_{s}+2g_{s1}\text{Re}(\beta_{1})$ are, respectively, the effective driving detunings of the cavity mode $a$ and the auxiliary cavity mode $a_{s}$, with $\text{Re}(\beta_{l=1,2})$ taking the real part of $\beta_{l}$. The parameter $G_{l}=g_{l}\alpha$ ($G_{s1}=g_{s1}\alpha_{s}$) is the linearized optomechanical-coupling strength between the cavity mode $a$ ($a_{s}$) and the $l$th mechanical mode (the mechanical mode $b_{1}$). In the steady-state case, the average values of the system operators can be obtained as
\begin{subequations}
\begin{align}
\alpha \equiv& \langle a\rangle_{\text{ss}}=\frac{-i\Omega}{\kappa +i\Delta_{c}^{\prime}}, \\
\alpha_{s} \equiv& \langle a_{s}\rangle_{\text{ss}}=\frac{-i\Omega_{s}}{\kappa_{s}+i\Delta_{s}^{\prime}}, \\
\beta_{1}\equiv&\langle b_{1}\rangle_{\text{ss}}=\frac{-ig_{1}\vert \alpha\vert^{2}-ig_{s1}\vert\alpha_{s}\vert^{2}}{\gamma_{1}+i\omega_{1}}, \\
\beta_{2}\equiv&\langle b_{2}\rangle_{\text{ss}}=\frac{-ig_{2}\vert \alpha\vert^{2}}{\gamma_{2}+i\omega_{2}}.
\end{align}
\end{subequations}
For convenience, in the following we assume that the steady-state values of $\alpha$ and $\alpha_{s}$ are real by choosing proper phases of the driving amplitudes $\Omega$ and $\Omega_{s}$; then the linearized optomechanical-coupling strengths $G_{l=1,2}$ and $G_{s1}$ are also real.

Based on Eq.~(\ref{linzedlang}), we can derive an approximate linearized Hamiltonian, which governs the dynamics of the system. To implement the cooling scheme, the system should work in the red-sideband-resonance regime, in which the rotating-wave approximation can be safely made. By discarding the noise terms, the linearized optomechanical Hamiltonian can be written as
\begin{eqnarray}
H_{\text{RWA}} &=&\Delta_{c}^{\prime}\delta a^{\dagger}\delta a+\Delta_{s}^{\prime}\delta a_{s}^{\dagger}\delta a_{s}+G_{s1}(\delta a_{s}\delta b_{1}^{\dagger}+\delta a_{s}^{\dagger}\delta b_{1}) \notag \\
&&+\sum_{l=1,2}\left[\omega_{l}\delta b_{l}^{\dagger}\delta b_{l}+G_{l}(\delta a\delta b_{l}^{\dagger}+\delta a^{\dagger}\delta b_{l})\right].  \label{Hamiltthree}
\end{eqnarray}

To clearly see the dark-mode effect in this $N$-type four-mode optomechanical system, we first consider the case where the auxiliary cavity is absent, i.e., $\Delta_{s}^{\prime}=0$ and $G_{s1}=0$. Then Hamiltonian~(\ref{Hamiltthree}) becomes
\begin{eqnarray}
H_{\text{RWA}}^{\prime}&=&\Delta_{c}^{\prime}\delta a^{\dagger}\delta a+ \sum_{l=1,2}\left[\omega_{l}\delta b_{l}^{\dagger}\delta b_{l}+G_{l}(\delta a\delta b_{l}^{\dagger}+\delta a^{\dagger}\delta
b_{l})\right]. \notag\\ \label{Hamilthrees1}
\end{eqnarray}
In this case, two hybrid mechanical modes $B_{+}$ and $B_{-}$ can be introduced as
\begin{subequations}
\begin{align}
B_{+}=&\frac{1}{\sqrt{G^{2}_{1}+G^{2}_{2}}}(G_{1}\delta b_{1}+G_{2}\delta b_{2}),\\
B_{-}=&\frac{1}{\sqrt{G^{2}_{1}+G^{2}_{2}}}(G_{2}\delta b_{1}-G_{1}\delta b_{2}).\label{BDmodedef}
\end{align}
\end{subequations}
Here, the new operators satisfy the bosonic commutation relations $[B_{+},B^{\dagger}_{+}]=1$ and $[B_{-},B^{\dagger}_{-}]=1$. By substituting the operators $B_{+}$ ($B^{\dag}_{+}$) and $B_{-}$ ($B^{\dag}_{-}$) into Eq.~(\ref{Hamilthrees1}), the Hamiltonian $H_{\text{RWA}}^{\prime}$ can be rewritten as
\begin{eqnarray}
H_{\text{RWA}}^{\prime}&=&\Delta_{c}^{\prime}\delta a^{\dagger}\delta a+\omega_{+}B_{+}^{\dagger}B_{+}+\omega_{-}B_{-}^{\dagger }B_{-}+G_{+}(\delta aB_{+}^{\dagger}+B_{+}\delta a^{\dagger}) \nonumber\\
&&+\zeta(B_{+}^{\dagger}B_{-}+B_{-}^{\dagger}B_{+}),\label{Hamitsdig}
\end{eqnarray}
where we introduce the resonance frequencies $\omega_{\pm}$ of the two hybrid modes and the coupling strengths $\zeta$ and $G_{+}$, which are defined by
\begin{subequations}
\label{omegazf}
\begin{align}
\omega_{+} =&\frac{\omega_{1}G^{2}_{1}+\omega_{2}G^{2}_{2}}{G^{2}_{1}+G^{2}_{2}},\\
\omega_{-} =&\frac{\omega_{1}G^{2}_{2}+\omega_{2}G^{2}_{1}}{G^{2}_{1}+G^{2}_{2}},\\
G_{+}=&\sqrt{G^{2}_{1}+G^{2}_{2}},\\
\zeta =&\frac{(\omega_{1}-\omega_{2})G_{1}G_{2}}{G^{2}_{1}+G^{2}_{2}}.
\end{align}
\end{subequations}
From Eqs.~(\ref{omegazf}) we can see that the two hybrid modes $B_{-}$ and $B_{+}$ are decoupled from each other ($\zeta=0$) when $\omega_{1}=\omega_{2}$. Moreover, the hybrid mode $B_{-}$ also decouples from the cavity mode $a$, which means that the hybrid mode $B_{-}$ becomes a dark mode. At this time, the ground-state cooling of the two mechanical modes is largely suppressed.

To break the dark-mode effect in this optomechanical system, we introduce an auxiliary cavity mode, which is coupled to the mechanical mode $b_{1}$ via the radiation-pressure interaction. By substituting operators $B_{+}$ ($B^{\dag}_{+}$) and $B_{-}$ ($B^{\dag}_{-}$) into Eq.~(\ref{Hamiltthree}), the Hamiltonian $H_{\text{RWA}}$ becomes
\begin{eqnarray}
H_{\text{RWA}}&=&\Delta_{c}^{\prime}\delta a^{\dagger}\delta a+\Delta_{s}^{\prime}\delta a_{s}^{\dagger}\delta a_{s}+\omega_{+}B_{+}^{\dagger}B_{+}+\omega_{-}B_{-}^{\dagger }B_{-}\nonumber\\
&&+\zeta(B_{+}^{\dagger}B_{-}+B_{-}^{\dagger}B_{+})+G_{+}(\delta aB_{+}^{\dagger}+B_{+}\delta a^{\dagger}) \nonumber\\
&&+G_{s+}(\delta a_{s}B_{+}^{\dagger}+\delta a_{s}^{\dagger}B_{+})+G_{s-}(\delta a_{s}B_{-}^{\dagger}+\delta a_{s}^{\dagger}B_{-}), \nonumber\\ \label{Hamitthreedig}
\end{eqnarray}
where we introduce two new coupling strengths $G_{s+}$ and $G_{s-}$,
\begin{subequations}
\label{gszffou}
\begin{align}
G_{s+}&=\frac{G_{s1}G_{1}}{\sqrt{G^{2}_{1}+G^{2}_{2}}},\\
G_{s-}&=\frac{G_{s1}G_{2}}{\sqrt{G^{2}_{1}+G^{2}_{2}}}.\\  \notag
\end{align}
\end{subequations}
The coupling configuration associated with the approximate Hamiltonian~(\ref{Hamitthreedig}) is described by Fig.~\ref{Fig1}(b). Here, the couplings are expressed in the representation of the mechanical hybrid modes. We can see from Eqs.~(\ref{gszffou}) that $G_{s\pm}>0$, i.e., the hybrid modes $B_{\pm}$ are always coupled with the auxiliary cavity mode $a_{s}$. Therefore, even if the hybrid mode $B_{-}$ is decoupled from both the intermediate cavity mode $a$ and the hybrid mode $B_{+}$, the ground-state cooling of the two mechanical modes can also be achieved via the cooling channel associated with the auxiliary cavity mode $a_{s}$.

For studying quantum cooling of the mechanical modes, we are interested in the steady-state properties of the system. Therefore, we should analyze the stability condition of this linearized system. To this end, we rewrite the linearized Langevin equations~(\ref{linzedlang}) as the contact form
\begin{eqnarray}
\mathbf{\dot{u}}(t)=\mathbf{Au}(t)+\mathbf{N}(t),\label{MatrixLeq}
\end{eqnarray}
where $\mathbf{u}(t)=[\delta a,\delta a_{s},\delta b_{1}, \delta b_{2},\delta a^{\dagger},\delta a_{s}^{\dagger},\delta b^{\dagger}_{1}, \delta b^{\dagger}_{2}]^{T}$ and $\mathbf{N}(t)=\sqrt{2}[\sqrt{\kappa}a_{\text{in}},\sqrt{\kappa_{s}}a_{s,\text{in}},\sqrt{\gamma_{1}}b_{1,\text{in}},\sqrt{\gamma_{2}}b_{2,\text{in}}
,\sqrt{\kappa}a^{\dagger}_{\text{in}},\sqrt{\kappa_{s}}a^{\dagger}_{s,\text{in}}, \sqrt{\gamma_{1}}\\b^{\dagger}_{1,\text{in}}, \sqrt{\gamma_{2}}b^{\dagger}_{2,\text{in}}]^{T}$ are, respectively, the fluctuation operator vector and noise operator vector with the matrix transpose notation ``$T$". The coefficient matrix is defined by $\mathbf{A}=\left(\begin{array}{cc}\mathbf{E} & \mathbf{F}\\
\mathbf{F}^{\ast} & \mathbf{E}^{\ast} \\\end{array}\right)$, where
\begin{eqnarray}
\mathbf{E}=-\left(
\begin{array}{cccc}
\kappa +i\Delta _{c}^{\prime} & 0 & iG_{1} & iG_{2} \\
0 &\kappa _{s}+i\Delta_{s}^{\prime} & iG_{s1} & 0 \\
iG_{1}^{\ast} & iG_{s1}^{\ast} &\gamma_{1}+i\omega_{1} & 0 \\
iG_{2}^{\ast} & 0 & 0 & \gamma_{2}+i\omega_{2}
\end{array}
\right),
\label{Ematrix}
\end{eqnarray}
and $\mathbf{F}$ is defined by the nonzero elements
$\mathbf{F}_{13}=-iG_{1}$, $\mathbf{F}_{14}=-iG_{2}$, $\mathbf{F}_{23}=-iG_{s1}$, $\mathbf{F}_{31}=-iG_{1}$, $\mathbf{F}_{32}=-iG_{s1}$, and $\mathbf{F}_{41}=-iG_{2}$.
The eigensystem of the coefficient matrix $\mathbf{A}$ determines the stability of the system. By using the Routh-Hurwitz criterion~\cite{Gradstein2014}, we can find the stability condition. In the following derivation, all the parameters used satisfy the stability conditions.

\section{Ground-state cooling of the two mechanical modes\label{cooling}}

In this section, we study the cooling performance of the two mechanical modes by calculating the final mean phonon numbers. The formal solution of the linearized Langevin equations~(\ref{MatrixLeq}) can be obtained as
\begin{equation}
\mathbf{u}(t) =\mathbf{M}(t) \mathbf{u}(0)+\int_{0}^{t}\mathbf{M}(t-s)\mathbf{N}(s)ds\label{formalsol},
\end{equation}
where we introduce the matrix $\mathbf{M}(t)=\exp(\mathbf{A}t)$. The final mean phonon numbers of the two mechanical modes can be calculated by solving the steady state of the system.

%%%%%%%%%%%%%%%%%%%%%%%%%%%%%%
\begin{figure}[tbp]
\centering
\includegraphics[width=0.48 \textwidth]{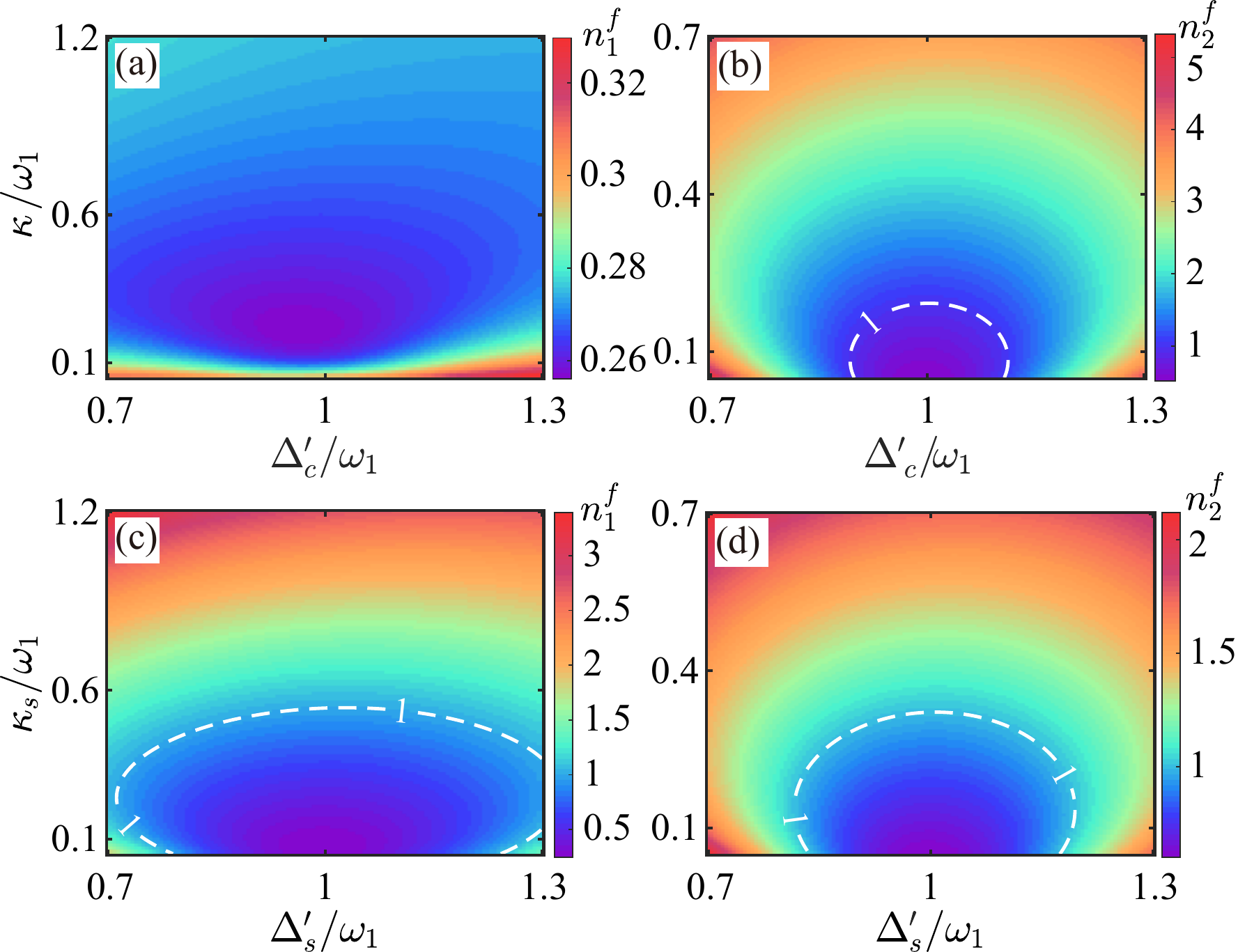}
\caption{Final mean phonon numbers (a) and (c) $n^{f}_{1}$ and (b) and (d) $n^{f}_{2}$ of the two mechanical modes versus the scaled driving detuning (a) and (b) $\Delta_{c}^{\prime}/\omega_{1}$ and (c) and (d) $\Delta_{s}^{\prime}/\omega_{1}$ and the scaled cavity-field decay rate (a) and (b) $\kappa/\omega_{1}$ and (c) and (d) $\kappa_{s}/\omega_{1}$. The parameters are (a) and (b) $\Delta_{s}^{\prime}/\omega_{1}=1$ and $\kappa_{s}/\omega_{1}=0.1$ and (c) and (d)  $\Delta_{c}^{\prime}/\omega_{1}=1$ and $\kappa/\omega_{1}=0.1$. The other parameters are $\omega_{2}/\omega_{1}=1$, $\gamma_{1}/\omega_{1}=\gamma_{2}/\omega_{1}=10^{-5}$, $G_{1}/\omega_{1}=G_{2}/\omega_{1}=0.05$, $G_{s1}/\omega_{1}=0.08$, and $\bar{n}_{1}=\bar{n}_{2}=1000$.}
\label{Fig2}
\end{figure}
%%%%%%%%%%%%%%%%%%%%%%%%%%%%%%

Based on Eq.~(\ref{formalsol}), we can obtain the final mean phonon numbers by solving the Lyapunov equation~\cite{Vitali2007}. To this end, we introduce the covariance matrix $\mathbf{V}$ of the system by defining the matrix elements as
\begin{equation}
\mathbf{V}_{ij}=\frac{1}{2}[\langle \mathbf{u}_{i}(\infty) \mathbf{u}_{j}(\infty ) \rangle +\langle \mathbf{u}_{j}( \infty) \mathbf{u}_{i}(\infty )\rangle], \hspace{0.5 cm}i,j=1,\ldots,8. \label{covariance}
\end{equation}
In the linearized optomechanical system, the covariance matrix $\mathbf{V}$ satisfies the Lyapunov equation
\begin{equation}
\mathbf{A}\mathbf{V}+\mathbf{V}\mathbf{A}^{T}=-\mathbf{Q}. \label{Lyapunov}
\end{equation}
Here, the matrix $\mathbf{Q}$ is defined by
\begin{equation}
\mathbf{Q}=\frac{1}{2}(\mathbf{C}+\mathbf{C}^{T}),\label{Qccstar}
\end{equation}
where $\mathbf{C}$ is the correlation matrix related to the noise operators. The matrix elements of $\mathbf{C}$ are defined by
\begin{eqnarray}
\langle \mathbf{N}_{k}(s)\mathbf{N}_{l}(s^{\prime})\rangle =\mathbf{C}_{k,l}\delta (s-s^{\prime }).
\end{eqnarray}
In this work  we consider the Markovian dissipation case. Then the constant matrix $\mathbf{C}$ can be obtained with the nonzero elements $\mathbf{C}_{15}=2\kappa$, $\mathbf{C}_{26}=2\kappa_{s}$, $\mathbf{C}_{37}=2\gamma_{1}(\bar{n}_{1}+1)$, $\mathbf{C}_{48}=2\gamma_{2}(\bar{n}_{2}+1)$, $\mathbf{C}_{73}=2\gamma _{1}\bar{n}_{1}$, and $\mathbf{C}_{84}=2\gamma_{2}\bar{n}_{2}$.
By solving the Lyapunov equation, we obtain the covariance matrix $\mathbf{V}$ defined in Eq.~(\ref{covariance}). Then the final mean phonon numbers of the two mechanical modes can be obtained as
\begin{subequations}
\label{finalexact}
\begin{align}
n^{f}_{1}=&\langle \delta b_{1}^{\dagger}\delta b_{1}\rangle=\mathbf{V}_{73}-\frac{1}{2},\\
n^{f}_{2}=&\langle \delta b_{2}^{\dagger}\delta b_{2}\rangle=\mathbf{V}_{84}-\frac{1}{2},
\end{align}
\end{subequations}
where $\mathbf{V}_{73}$ and $\mathbf{V}_{84}$ are the matrix elements of the covariance matrix $\mathbf{V}$.

Here, we first consider the cooling performance in the two-degenerate-resonator case ($\omega_{1}=\omega_{2}$). In Fig.~\ref{Fig2} we plot the final mean phonon numbers $n_{1}^{f}$ and $n_{2}^{f}$ as functions of the scaled driving detuning $\Delta_{c}^{\prime}/\omega_{1}$ ($\Delta_{s}^{\prime}/\omega_{1}$) and the scaled cavity-field decay rate $\kappa/\omega_{1}$ ($\kappa_{s}/\omega_{1}$). To better analyze the influence of the driving detuning and the sideband-resolution condition on the cooling performance, here we choose the mechanical frequency $\omega_{1}$ as the scale unit. In Figs.~\ref{Fig2}(a) and~\ref{Fig2}(b), we can see that ground-state cooling of the two mechanical modes can be realized in the resolved-sideband limit ($\kappa/\omega_{1}\ll1$) and around $\Delta_{c}^{\prime}/\omega_{1}\sim1$. Moreover, the cooling efficiency of the first mechanical mode is much better than that of the second one even in unresolved-sideband limit ($\kappa/\omega_{1}\geq1$). This is because the first mechanical mode is simultaneously connected to two cooling channels and the coupling strength between the auxiliary cavity mode and the first mechanical mode is large enough ($G_{s1}/\omega_{1}=0.08$). For a given decay rate $\kappa/\omega_{1}$, the optimal driving detuning is about $\Delta_{c}^{\prime}/\omega_{1}=1$, which corresponds to the red-sideband resonance. In Figs.~\ref{Fig2}(c) and~\ref{Fig2}(d), we can see that the ground-state cooling of the two mechanical modes can be realized in the resolved-sideband limit ($\kappa_{s}/\omega_{1}\ll1$), and the cooling performance is the best at the optimal driving detuning $\Delta_{s}^{\prime}/\omega_{1}\approx1$. These results are consistent with the sideband cooling results in typical optomechanical systems~\cite{Wilson-Rae2007,Marquardt2007,Genes2008a}. In addition, we perform numerical calculations with several sets of parameters when $\Delta_{s}^{\prime}/\omega_{1}\gg1$ or $\kappa_{s}/\omega_{1}\gg1$. We find that though these mechanical modes can be cooled significantly, but they cannot be cooled to the ground state in the $N$-type optomechanical system.

%%%%%%%%%%%%%%%%%%%%%%%%%%%%%
\begin{figure}[tbp]
\centering
\includegraphics[width=0.47 \textwidth]{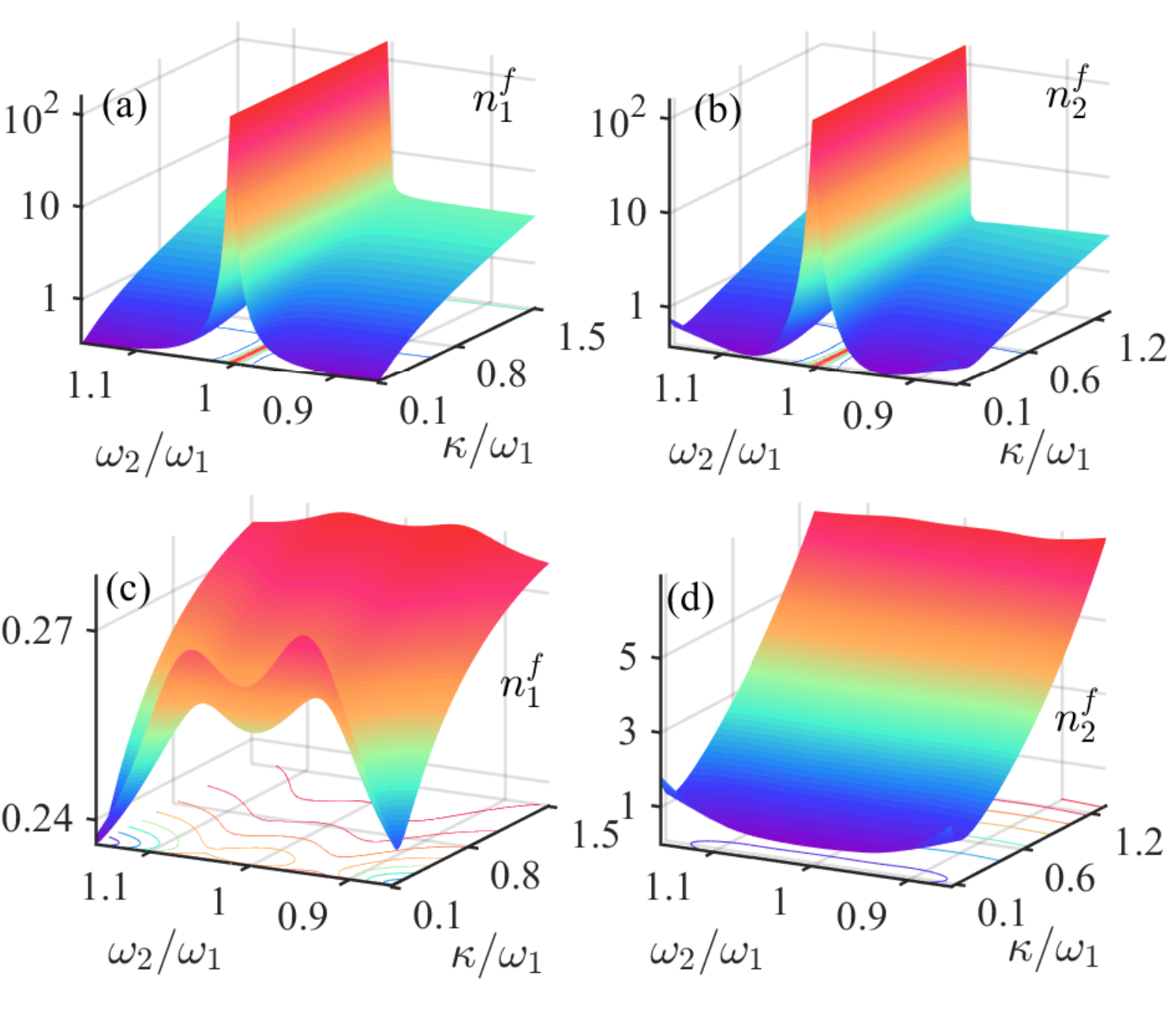}
\caption{Final mean phonon numbers $n^{f}_{1}$ and $n^{f}_{2}$ versus the frequency ratio $\omega_{2}/\omega_{1}$ and the scaled cavity-field decay rate $\kappa/\omega_{1}$ in  the cases of (a) and (b) dark-mode unbreaking ($G_{s1}/\omega_{1}=0$) and (c) and (d)  dark-mode breaking ($G_{s1}/\omega_{1}\neq0$). The other parameters are $\Delta_{c}^{\prime}=\Delta_{s}^{\prime}=\omega_{1}$, $\kappa_{s}/\omega_{1}=0.1$, $G_{1}/\omega_{1}=G_{2}/\omega_{1}=0.05$, $G_{s1}/\omega_{1}=0.08$, $\gamma_{1}/\omega_{1}=\gamma_{2}/\omega_{1}=10^{-5}$, and $\bar{n}_{1}=\bar{n}_{2}=1000$.}
\label{Fig3}
\end{figure}
%%%%%%%%%%%%%%%%%%%%%%%%%%%%%%

Next we analyze the influence of the frequency mismatch between the two mechanical modes on the cooling efficiency. In Figs.~\ref{Fig3}(a) and~\ref{Fig3}(b) we plot the phonon numbers $n_{1}^{f}$ and $n_{2}^{f}$ versus the frequency ratio $\omega_{2}/\omega_{1}$ and the scaled cavity-field decay rate $\kappa/\omega_{1}$ in the absence of the auxiliary cavity mode ($G_{s1}/\omega_{1}=0$). Here, we see that the final mean phonon numbers in the two mechanical modes cannot be efficiently decreased around $\omega_{2}=\omega_{1}$, which means that the two mechanical modes cannot be cooled to their ground states when their frequencies are degenerate or nearly degenerate (in a finite-detuning window). This phenomenon can be explained according to the dark-mode effect. In the degenerate-resonator case ($\omega_{1}=\omega_{2}$), a bright mode and a dark mode are formed in this optomechanical system. Physically, the two mechanical modes have an obvious spectral overlap and become effectively degenerate in the presence of dissipation, thus the dark-mode effect works in the near-degenerate case. The dark mode decouples from both the cavity mode and the bright mode, so the phonon excitations stored in the dark mode cannot be extracted through the optomechanical-cooling channel~\cite{Lai2020}. However, the dark-mode effect disappears when the two mechanical modes are far-off-resonance, thus achieving ground-state cooling. In Figs.~\ref{Fig3}(c) and~\ref{Fig3}(d)  we plot the phonon numbers $n_{1}^{f}$ and $n_{2}^{f}$ versus the frequency ratio $\omega_{2}/\omega_{1}$ and the scaled decay rate $\kappa/\omega_{1}$, when the auxiliary cavity field is present ($G_{s1}/\omega_{1}=0.08$). Here the ground-state cooling of the two mechanical modes can be achieved ($n^{f}_{1,2}\ll1$) when the system works in the resolved-sideband regime ($\kappa\ll\omega_{1}$). Here the first mechanical resonator has a better cooling efficiency  ($n^{f}_{1}<n^{f}_{2}$) because it is connected to two cooling channels at the same time.

%%%%%%%%%%%%%%%%%%%%%%%%%%%%%%
\begin{figure}[tbp]
\centering
\includegraphics[width=0.47 \textwidth]{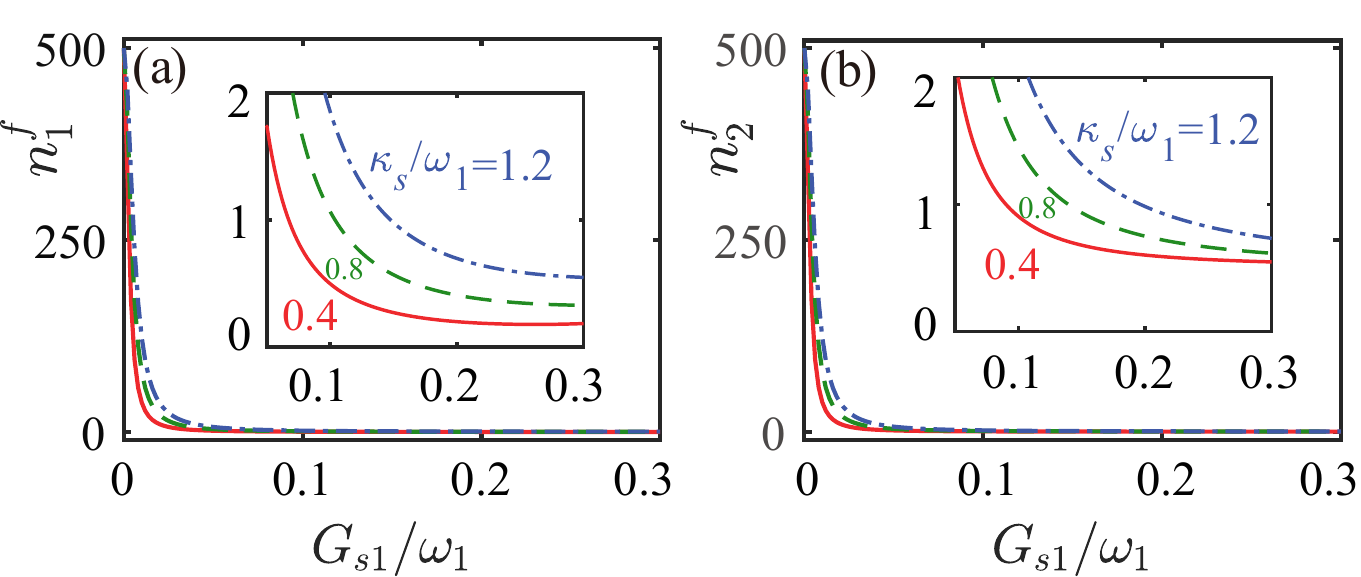}
\caption{Final mean phonon numbers (a) $n^{f}_{1}$ and (b) $n^{f}_{2}$ as functions of $G_{s1}/\omega_{1}$ when the scaled cavity-field decay rate takes various values $\kappa_{s}/\omega_{1}=0.4$, $0.8$, and $1.2$. The insets show close-ups of the final mean phonon numbers as functions of $G_{s1}/\omega_{1}$, with the horizontal axis ranging from $0.05$ to $0.3$. The other parameters are $\omega_{2}/\omega_{1}=1$, $\Delta_{c}^{\prime}/\omega_{1}=\Delta_{s}^{\prime}/\omega_{1}=1$, $\kappa/\omega_{1}=0.1$, $\gamma_{1}/\omega_{1}=\gamma_{2}/\omega_{1}=10^{-5}$, $G_{1}/\omega_{1}=G_{2}/\omega_{1}=0.05$, and $\bar{n}_{1}=\bar{n}_{2}=1000$.}
\label{Fig4}
\end{figure}
%%%%%%%%%%%%%%%%%%%%%%%%%%%%%%

Since the auxiliary cavity mode not only provides the direct channel to extract the thermal excitations from the mechanical mode $b_{1}$, but also provides a cooling channel to extract the thermal excitations from the mechanical mode $b_{2}$, the coupling strength between the auxiliary cavity mode $a_{s}$ and the mechanical mode $b_{1}$ is an important factor to the cooling efficiency. To see this effect more clearly, in Figs.~\ref{Fig4}(a) and~\ref{Fig4}(b) we plot the phonon numbers $n_{1}^{f}$ and $n_{2}^{f}$ as functions of the linearized optomechanical-coupling strength $G_{s1}$ between the auxiliary cavity mode $a_{s}$ and the mechanical mode $b_{1}$ when the scaled cavity-field decay rate $\kappa_{s}/\omega_{1}$ takes various values. Here, we see that the phonon numbers decrease with the increase of the coupling strength $G_{s1}$, which means that the increase of the coupling strength $G_{s1}$ is beneficial to the ground-state cooling of the two mechanical resonators. Moreover, the final mean phonon numbers are smaller for smaller values of the decay rate $\kappa_{s1}/\omega_{1}$ for a certain parameter range, which is consistent with the analyses of the cooling efficiency on the sideband-resolution condition~\cite{Wilson-Rae2007,Marquardt2007,Genes2008a}.In addition, we can see from Fig.~\ref{Fig4} that the ground-state cooling can be realized in the $N$-type optomechanical system and the cooling limit cannot be broken compared with the single-resonator optomechanical system under the same parameters.

\section{Ground-state cooling and universal dark-mode-breaking conditions in the four-mode optomechanical system \label{Universal}}

In previous sections we have shown that, by introducing an optomechanical coupling between the auxiliary cavity mode $a_{s}$ and the mechanical mode $b_{1}$, the dark mode in this system can be broken and then ground-state cooling of the two mechanical modes can be realized. However, in practice, the diverse interactions among these degrees of freedom in this system are more complicated~\cite{Ludwig2013}, so it is an interesting topic to study the universal condition for breaking the dark-mode effect in a more general four-mode optomechanical system. In this section, we analyze the parameter conditions under which the dark-mode effect works and study how to break the dark mode by controlling the couplings in the four-mode optomechanical system. We also analyze the interference effect by studying the energy-level transition of this four-mode optomechanical system.

\subsection{Ground-state cooling in the general four-mode optomechanical system \label{coolingnet}}

%%%%%%%%%%%%%%%%%%%%%%%%%%%%%%
\begin{figure}[tbp]
\center
\includegraphics[width=0.47 \textwidth]{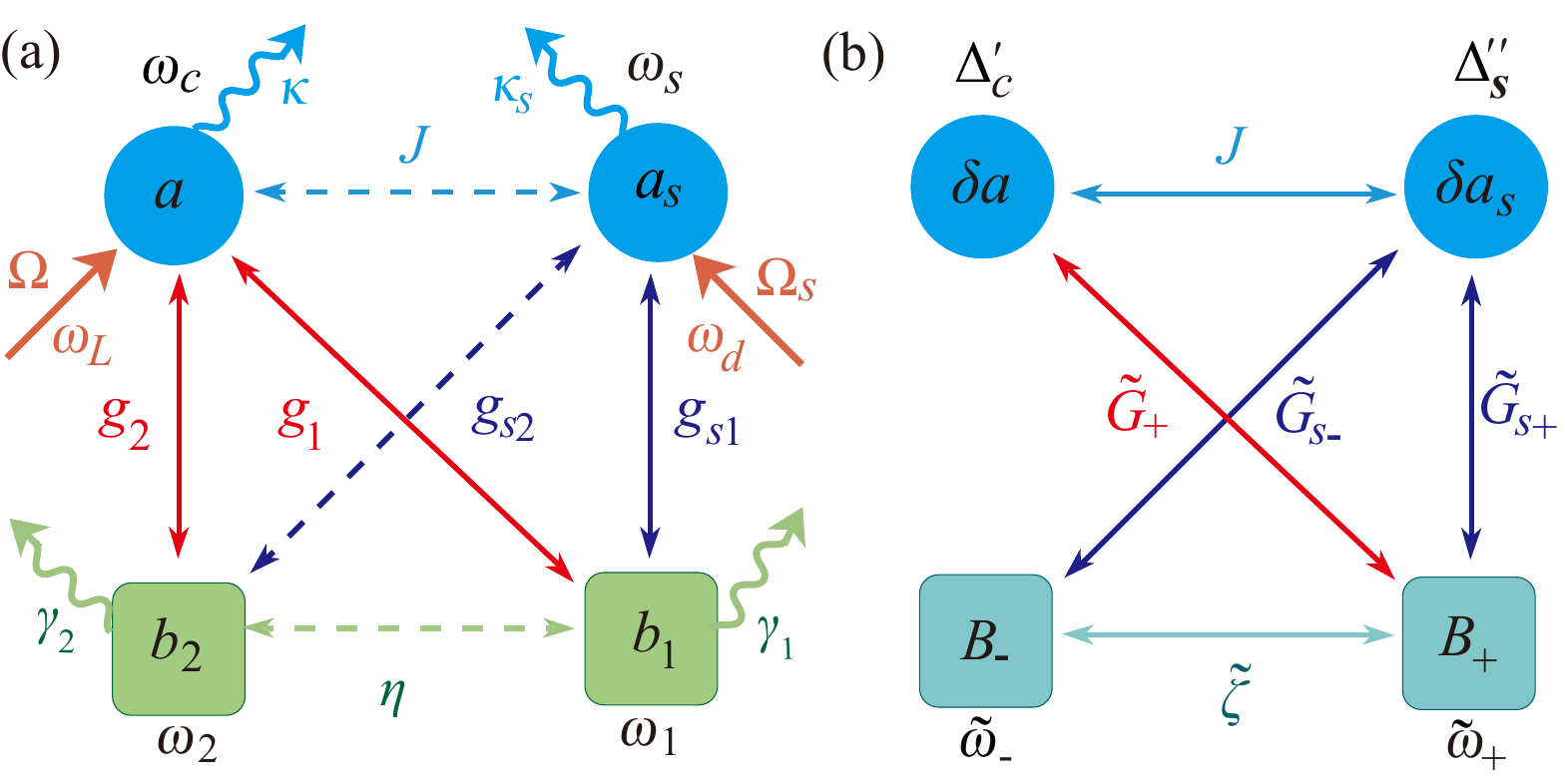}
\caption{(a) Schematic of the network-coupled four-mode optomechanical system consisting of two cavity modes and two mechanical modes. In addition to the couplings and notations introduced in Fig.~\ref{Fig1}(a), here we introduce three new couplings marked by dashed lines: the phonon- and photon-hopping interactions with the coupling strength $\eta$ and $J$, respectively, and the optomechanical-coupling strength $g_{s2}$ between the auxiliary cavity mode $a_{s}$ and the second mechanical mode $b_{2}$. (b) Coupling configuration associated with the approximate linearized Hamiltonian~(\ref{Hamitfulldig}). Here we add a tilde to the couplings and notation introduced in Fig.~\ref{Fig1}(b). We also introduce the photon-hopping strength $J$ between the two cavity modes.}
\label{Fig5}
\end{figure}
%%%%%%%%%%%%%%%%%%%%%%%%%%%%%%

We now consider a network-coupled four-mode optomechanical system consisting of an intermediate coupling cavity mode and an auxiliary cavity mode, which are both coupled to two mechanical modes via the radiation-pressure interaction. Here, the two cavity (mechanical) modes are coupled to each other via a photon-hopping (phonon-hopping) interaction, as shown in Fig.~\ref{Fig5}(a). In a rotating frame defined by the operator $\exp[-i(\omega _{L}a^{\dagger}a+\omega_{d}a_{s}^{\dagger}a_{s})t]$ under $\omega _{L}=\omega_{d}$, the transformed Hamiltonian becomes
\begin{eqnarray}
H_{I}&=&\Delta_{c}a^{\dagger}a+\Delta_{s}a_{s}^{\dagger}a_{s}+J(a^{\dagger}a_{s}+a_{s}^{\dagger}a)+\eta(b_{1}^{\dagger}b_{2}+b_{2}^{\dagger}b_{1})\nonumber\\
&&+\sum_{l=1,2}[\omega_{l}b_{l}^{\dagger}b_{l}+g_{l}a^{\dagger}a(b_{l}^{\dagger}+b_{l})+g_{sl}a_{s}^{\dagger}a_{s}(b_{l}^{\dagger}+b_{l})]\nonumber\\
&&+(\Omega a^{\dagger}+\Omega_{s}a_{s}^{\dagger}+\mathrm{H.c.}), \label{Hamitfull}
\end{eqnarray}
where some operators and variables have been defined in Eq.~(\ref{Hamit1}). We also introduce the $g_{s2}$ coupling term, the $J$ coupling term, and the $\eta$ coupling term, which correspond to the optomechanical coupling between modes $a_{s}$ and $b_{2}$, the photon-hopping coupling, and the phonon-hopping coupling, respectively.

%%%%%%%%%%%%%%%%%%%%%%%%%%%%%%
\begin{figure*}[tbp]
\center
\includegraphics[width=0.85 \textwidth]{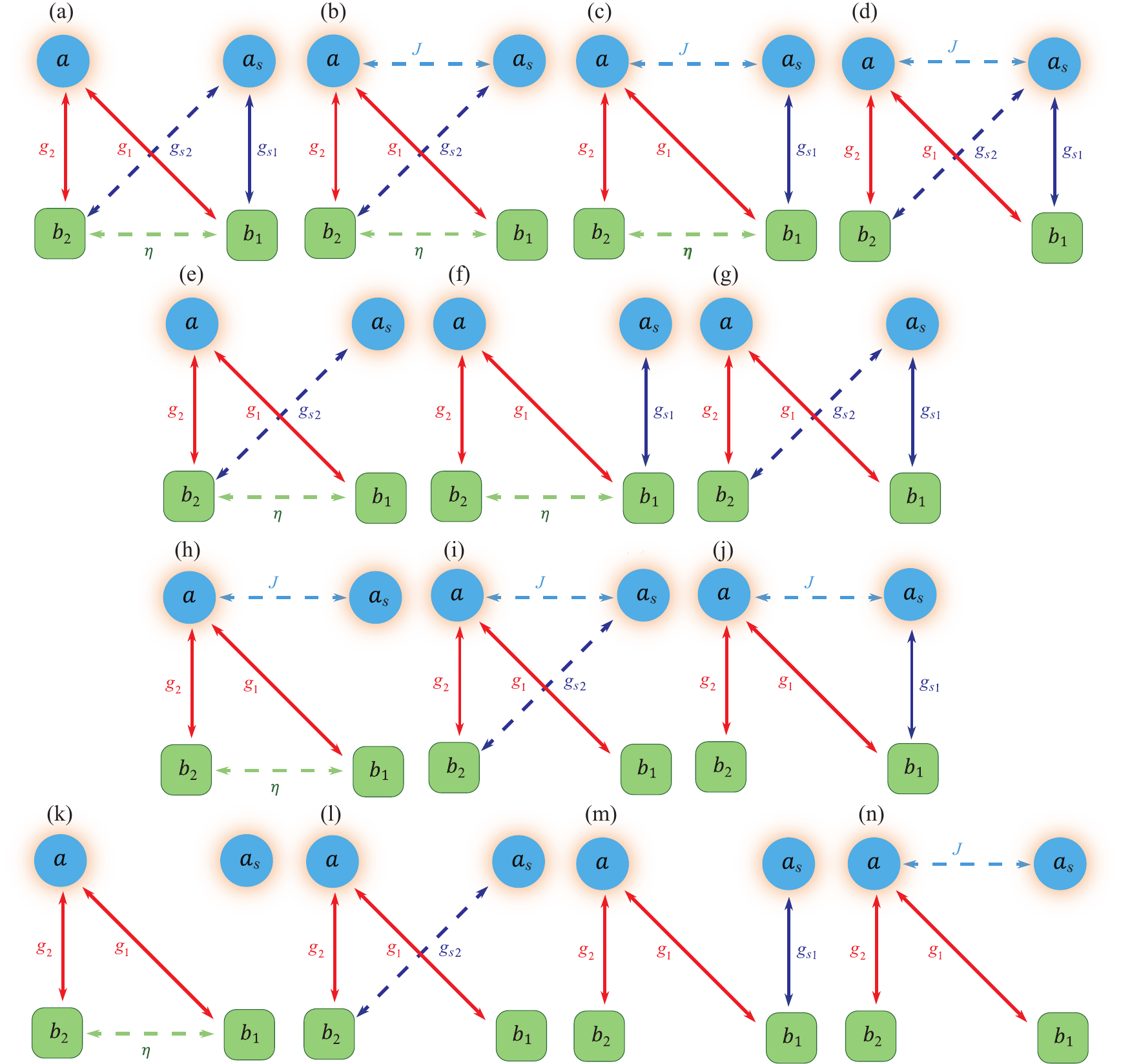}
\caption{Fourteen coupling configurations of the network-coupled four-mode optomechanical system, where $J$, $g_{l}$  ($g_{sl}$) (for $l=1,2$), and $\eta$ are the photon-hopping coupling strength, the optomechanical-coupling strength between the intermediate (auxiliary) cavity mode and the $l$th mechanical mode, and the phonon-hopping coupling strength, respectively. Here, the two optomechanical couplings between the cavity mode $a$ and the two mechanical modes $b_{1}$ and $b_{2}$ are kept  and the other couplings can be closed on demand. Then there are $14$ different coupling configurations. Concretely, the one-coupling-closed cases include the following: (a) The coupling channel $J$ is closed ($J=0$), (b) the coupling channel $g_{s1}$ is closed ($g_{s1}=0$), (c) the coupling channel $g_{s2}$ is closed ($g_{s2}=0$), and (d) the coupling channel $\eta$ is closed ($\eta=0$). The two-coupling-closed cases include the following: (e) The coupling channels $J$ and $g_{s1}$ are closed ($J=g_{s1}=0$), (f) the coupling channels $J$ and $g_{s2}$ are closed ($J=g_{s2}=0$), (g) the coupling channels $J$ and $\eta$ are closed ($J=\eta=0$), (h) the coupling channels $g_{s1}$ and $g_{s2}$ are closed ($g_{s1}=g_{s2}=0$), (i) the coupling channels $g_{s1}$ and $\eta$ are closed ($g_{s1}=\eta=0$), and (j) the coupling channels $g_{s2}$ and $\eta$ are closed ($g_{s2}=\eta=0$). In addition, the cases in which three couplings are closed include the following: (k) The coupling channels $J$, $g_{s1}$, and $g_{s2}$ are closed ($J=g_{s1}=g_{s2}=0$), (l) the coupling channels $J$, $g_{s1}$, and $\eta$ are closed ($J=g_{s1}=\eta=0$), (m) the coupling channels $J$, $g_{s2}$, and $\eta$ are closed ($J=g_{s2}=\eta=0$), and (n) the coupling channels $g_{s1}$, $g_{s2}$, and $\eta$ are closed ($g_{s1}=g_{s2}=\eta=0$). We point out that the single-photon optomechanical-coupling strengths $g_{1}$, $g_{2}$, $g_{s1}$, and $g_{s2}$ are related to the linearized optomechanical-coupling strengths $G_{1}$, $G_{2}$, $G_{s1}$, and $G_{s2}$ in the four-mode optomechanical system.}
\label{Fig6}
\end{figure*}
%%%%%%%%%%%%%%%%%%%%%%%%%%%%%%

%%%%%%%%%%%%%%%%%%%%%%%%%%%%%%
\begin{figure*}[tbp]
\centering
\includegraphics[width=0.85 \textwidth]{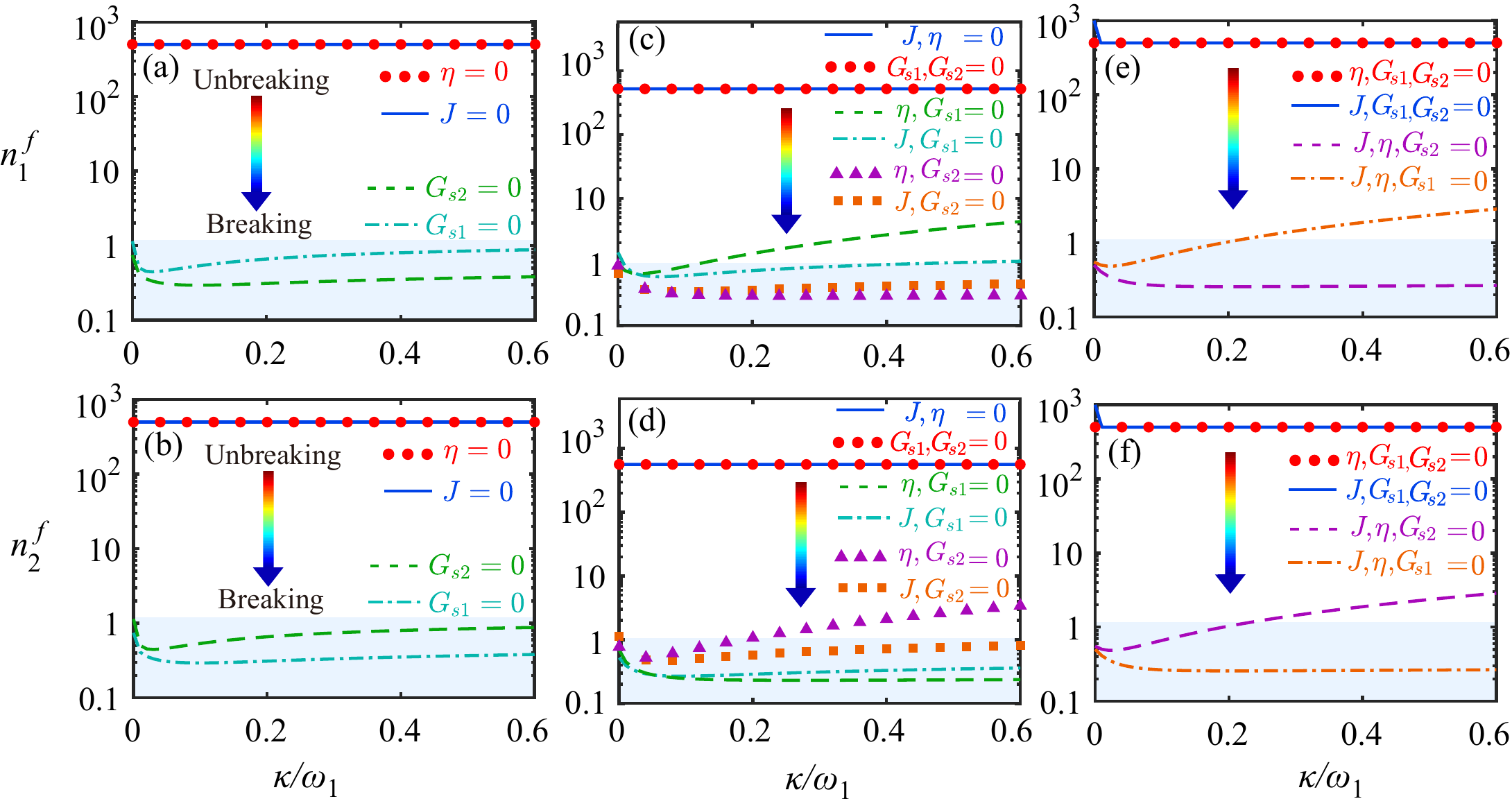}
\caption{Final mean phonon numbers $n^{f}_{1}$ and $n^{f}_{2}$ versus the scaled decay rate $\kappa/\omega_{1}$ in various coupling configurations: (a) and (b) cases where one coupling is closed, i.e., $J=0$ (blue solid curves), $\eta=0$ (red circles), $G_{s1}=0$ (cyan dash-dotted curves), and $G_{s2}=0$ (green dashed curves); (c) and (d) two-coupling-closed cases, i.e., $J=\eta=0$ (blue solid curves), $G_{s1}=G_{s2}=0$ (red circles), $\eta=G_{s1}=0$ (green dashed curves), $J=G_{s1}=0$ (cyan dash-dotted  curves), $J=G_{s2}=0$ (brown squares), and $\eta=G_{s2}=0$ (purple triangles);  and (e) and (f) three-coupling-closed cases, i.e., $J=G_{s1}=G_{s2}=0$ (blue solid curves), $\eta=G_{s1}=G_{s2}=0$ (red circles), $J=\eta=G_{s1}=0$ (brown dash-dotted curves), and $J=\eta=G_{s2}=0$ (brown dashed curves). The other parameters are $\omega_{2}/\omega_{1}=1$, $\gamma_{1}/\omega_{1}=\gamma_{2}/\omega_{1}=10^{-5}$, $\kappa/\omega_{1}=\kappa_{s}/\omega_{1}=0.1$, $J/\omega_{1}=\eta/\omega_{1}=0.03$, $\Delta_{c}^{\prime}/\omega_{1}=\Delta_{s}^{\prime\prime}/\omega_{1}=1$, $G_{1}/\omega_{1}=G_{2}/\omega_{1}=0.05$, $G_{s1}/\omega_{1}=G_{s2}/\omega_{1}=0.08$, and $\bar{n}_{1}=\bar{n}_{2}=1000$. Note that the values of $J$, $\eta$, $G_{s1}$, and $G_{s2}$ presented here work when the corresponding coupling channels are open.}
\label{Fig7}
\end{figure*}
%%%%%%%%%%%%%%%%%%%%%%%%%%%%%%

Based on the Hamiltonian~(\ref{Hamitfull}), we can obtain the Langevin equations by adding the decay and noise terms to the Heisenberg equations. Following a linearization procedure similar to the one performed in Sec.~\ref{model}, we obtain the linearized Langevin equations
\small
\begin{subequations}
\label{linzedlangfull}
\begin{align}
\delta \dot{a}=&-(\kappa+i\Delta_{c}^{\prime})\delta a-iJ\delta a_{s}-i\sum_{l=1,2}[G_{l}(\delta b_{l}+\delta b_{l}^{\dagger})]
+\sqrt{2\kappa }a_{\textrm{in}},  \\
\delta \dot{a}_{s} =&-(\kappa_{s}+i\Delta_{s}^{\prime\prime})\delta a_{s}-iJ\delta a-i\sum_{l=1,2}[G_{sl}(\delta b_{l}+\delta b_{l}^{\dagger})]+\sqrt{2\kappa _{s}}a_{s,\textrm{in}}, \\
\delta \dot{b}_{1}=&-(i\omega _{1}+\gamma _{1}) \delta b_{1}-iG_{1}^{\ast}\delta a-iG_{1}\delta a^{\dagger}-iG_{s1}^{\ast}\delta a_{s}-iG_{s1}\delta a_{s}^{\dagger}\nonumber \\
&-i\eta \delta b_{2}+\sqrt{2\gamma_{1}}b_{1,\textrm{in}}, \\
\delta \dot{b}_{2}=&-( i\omega_{2}+\gamma_{2})\delta b_{2}-iG_{2}^{\ast}\delta a-iG_{2}\delta a^{\dagger}-iG_{s2}^{\ast }\delta a_{s}-iG_{s2}\delta a_{s}^{\dagger}\nonumber \\
&-i\eta \delta b_{1}+\sqrt{2\gamma_{2}}b_{2,\textrm{in}},
\end{align}
\end{subequations}
\normalsize
where $\Delta_{s}^{\prime\prime}=\Delta_{s}+2g_{s1}\text{Re}(\beta_{1}) +2g_{s2}\text{Re}(\beta_{2})$ is the renormalized driving detuning of the auxiliary cavity mode $a_{s}$. It should be pointed out that the parameter $\Delta_{c}^{\prime}$ and the linearized optomechanical-coupling strengths $G_{l}$ and $G_{sl}$ for $l=1,2$ have the same definition as those defined in Sec.~\ref{model}. However, the coherent displacements of the steady state $\alpha$, $\alpha_{s}$, $\beta_{1}$, and $\beta_{2}$ in the network-coupled four-mode optomechanical system should be replaced by the relations
\begin{subequations}
\begin{align}
\alpha \equiv& \langle a\rangle_{\text{ss}}=\frac{-i(J\alpha_{s}+\Omega)}{\kappa +i\Delta_{c}^{\prime}}, \\
\alpha_{s} \equiv& \langle a_{s}\rangle_{\text{ss}}=\frac{-i(J\alpha +\Omega_{s})}{\kappa_{s}+i\Delta_{s}^{\prime\prime}}, \\
\beta_{1}\equiv&\langle b_{1}\rangle_{\text{ss}}=\frac{-i(\eta\beta_{2}+g_{1}\vert \alpha\vert^{2}+g_{s1}\vert\alpha_{s}\vert^{2})}{\gamma_{1}+i\omega_{1}}, \\
\beta_{2}\equiv&\langle b_{2}\rangle_{\text{ss}}=\frac{-i(\eta\beta_{1}+g_{2}\vert \alpha\vert^{2}+g_{s2}\vert\alpha_{s}\vert^{2})}{\gamma_{2}+i\omega_{2}}.
\end{align}
\end{subequations}

In the study of quantum cooling of the mechanical modes, the linearized Langevin equations~(\ref{linzedlangfull}) can be written as a compact form $\mathbf{\dot{u}}(t)=\mathbf{A^{\prime}u}(t)+\mathbf{N}(t)$, where the form of the fluctuation operator vector $\mathbf{u}(t)$ and the noise operator vector $\mathbf{N}(t)$ is the same as those defined in Sec.~\ref{cooling}, while the coefficient matrix is given by
$\mathbf{A'}=\left(\begin{array}{cc}\mathbf{E'} & \mathbf{F'}\\
\mathbf{F'}^{\ast} & \mathbf{E'}^{\ast} \\\end{array}\right)$, where
\begin{equation}
\label{coefficientnet}
\mathbf{E'}=\left(
\begin{array}{cccc}
-(\kappa+i\Delta_{c}^{\prime}) & -iJ & -iG_{1} & -iG_{2}  \\
-iJ & -(i\Delta_{s}^{\prime\prime}+\kappa_{s}) & -iG_{s1} & -iG_{s2}  \\
-iG_{1}^{\ast} & -iG_{s1}^{\ast} & -(i\omega_{1}+\gamma_{1}) & -i\eta  \\
-iG_{2}^{\ast} & -iG_{s2}^{\ast} & -i\eta & -(i\omega_{2}+\gamma_{2})
\end{array}
\right),
\end{equation}
and $\mathbf{F'}$ is defined by the nonzero elements
$\mathbf{F}_{13}=-iG_{1}$, $\mathbf{F}_{14}=-iG_{2}$, $\mathbf{F}_{23}=-iG_{s1}$, $\mathbf{F}_{24}=-iG_{s2}$, $\mathbf{F}_{31}=-iG_{1}$, $\mathbf{F}_{32}=-iG_{s1}$, $\mathbf{F}_{41}=-iG_{2}$, and $\mathbf{F}_{42}=-iG_{s2}$.

Following the same procedure as that performed in Sec.~\ref{cooling}, we can also obtain the steady-state expression of the covariance matrix $\mathbf{V^{\prime}}$, which is defined by $\mathbf{A^{\prime}}\mathbf{V^{\prime}}+\mathbf{V^{\prime}}\mathbf{A^{\prime}}^{T}=-\mathbf{Q}$, where $\mathbf{Q}$ is given by Eq.~(\ref{Qccstar}). Then the final mean phonon numbers in the two mechanical resonators can be obtained.

To clearly analyze the influence of these couplings in the four-mode optomechanical system on the ground-state cooling, below we consider various cases of different coupling configurations, as shown in Fig.~\ref{Fig6}. To analyze the dark-mode effect in this system when the frequencies of the two mechanical modes are degenerate, we consider the case where the two coupling channels $g_{1}$ and $g_{2}$ always exist. Then we study various cases of coupling configurations by controlling the four coupling channels $J$, $g_{s1}$, $g_{s2}$, and $\eta$. In Figs.~\ref{Fig6}(a)-\ref{Fig6}(d) and Figs.~\ref{Fig6}(e)-\ref{Fig6}(j), we show that one or two coupling channels of $J$, $g_{s1}$, $g_{s2}$, and $\eta$ are closed, respectively. In Figs.~\ref{Fig6}(k)-\ref{Fig6}(n), three of the four coupling channels $J$, $g_{s1}$, $g_{s2}$, and $\eta$ are closed. Therefore, when the coupling channels $g_{1}$ and $g_{2}$ still exist, there are $14$ cases of coupling configurations, as shown in Fig.~\ref{Fig6}.

Corresponding to the $14$ cases depicted in Fig.~\ref{Fig6}, we plot the final mean phonon numbers $n^{f}_{1}$ and $n^{f}_{2}$ as functions of the scaled decay rate $\kappa/\omega_{1}$ in Fig.~\ref{Fig7}. We can see from Figs.~\ref{Fig7}(a) and~\ref{Fig7}(b) that ground state cooling of the two mechanical modes can (cannot) be realized in the cases of $G_{s1}=0$ or $G_{s2}=0$ ($J=0$ or $\eta=0$), which implies that the dark mode can (cannot) be broken. In Figs.~\ref{Fig7}(c) and~\ref{Fig7}(d) we plot the phonon numbers $n^{f}_{1}$ and $n^{f}_{2}$ as functions of $\kappa/\omega_{1}$ when two of the four coupling channels ($J$, $g_{s1}$, $g_{s2}$, and $\eta$) are closed. Based on the cooling performance, we know that the dark mode cannot be broken when the coupling channels $J=\eta=0$ or $G_{s1}=G_{s2}=0$. In the four cases $J=G_{s1}=0$, $J=G_{s2}=0$, $\eta=G_{s1}=0$, and $\eta=G_{s2}=0$, the dark mode can be broken. In Figs.~\ref{Fig7}(e) and~\ref{Fig7}(f)  we also plot the phonon numbers $n^{f}_{1}$ and $n^{f}_{2}$ versus $\kappa/\omega_{1}$ when three of the four coupling channels are closed, corresponding to the cases shown in Figs.~\ref{Fig6}(k)-\ref{Fig6}(n). Here  we can see that the dark mode cannot be broken in the cases of $J=G_{s1}=G_{s2}=0$ or $\eta=G_{s1}=G_{s2}=0$. However, in the two cases of $J=\eta=G_{s1}=0$ or $J=\eta=G_{s2}=0$, the dark mode can be broken.

Based on the above discussion, we find that $G_{s1}$ and $G_{s2}$ play an important role in the breaking of the dark mode in this system. Only when one of $G_{s1}$ and $G_{s2}$ is closed, can the dark mode be broken. When neither or both $G_{s1}$ and $G_{s2}$ are closed, the dark mode cannot be broken. In particular, the breaking of the dark mode is independent of both the photon- and phonon-coupling channels.

%%%%%%%%%%%%%%%%%%%%%%%%%%%%%%
\begin{figure}[tbp]
\center
\includegraphics[width=0.48 \textwidth]{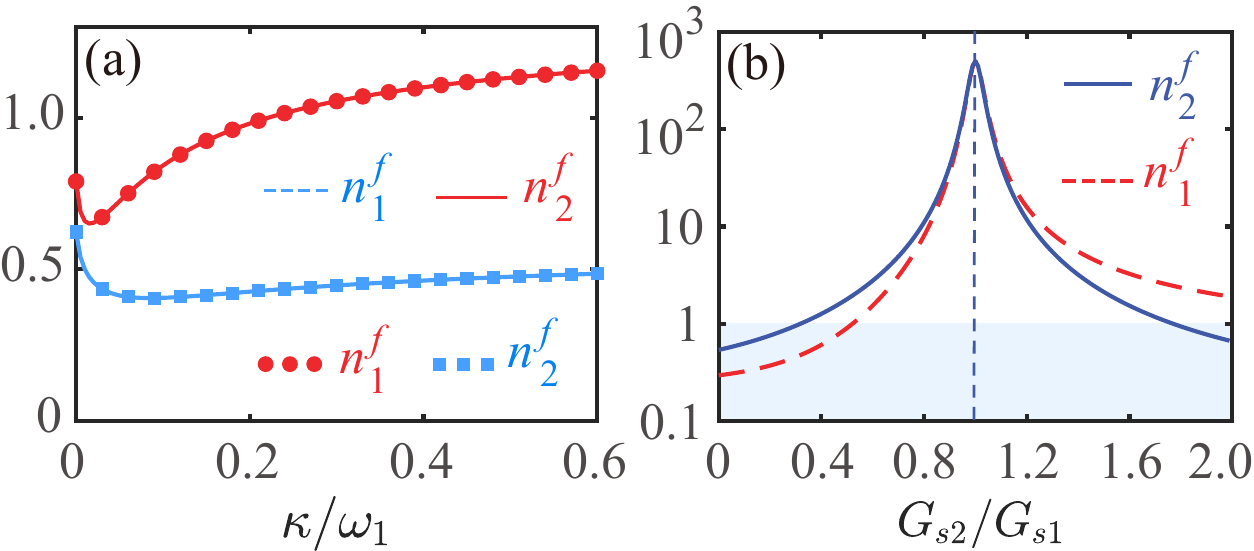}
\caption{(a) Final mean phonon numbers $n^{f}_{1}$ and $n^{f}_{2}$ versus the scaled decay rate $\kappa/\omega_{1}$ in the cases of $G_{s2}/\omega_{1}=4G_{s1}/\omega_{1}=0.08$ (blue dashed curve and red solid curve, respectively) and $G_{s1}/\omega_{1}=4G_{s2}/\omega_{1}=0.08$ (red circles and blue squares, respectively). (b) Final mean phonon numbers $n^{f}_{1}$ (blue solid curves) and $n^{f}_{2}$ (red dashed curves) versus the ratio $G_{s2}/G_{s1}$ when the linearized optomechanical-coupling strength $G_{s1}/\omega_{1}=0.08$. The other parameters used in this system are $\omega_{2}/\omega_{1}=1$, $\gamma_{1}/\omega_{1}=\gamma_{2}/\omega_{1}=10^{-5}$, $\kappa/\omega_{1}=\kappa_{s}/\omega_{1}=0.1$, $J/\omega_{1}=\eta/\omega_{1}=0.03$, $\Delta_{c}^{\prime}/\omega_{1}=\Delta_{s}^{\prime\prime}/\omega_{1}=1$, $G_{1}/\omega_{1}=G_{2}/\omega_{1}=0.05$, and $\bar{n}_{1}=\bar{n}_{2}=1000$.}
\label{Fig8}
\end{figure}
%%%%%%%%%%%%%%%%%%%%%%%%%%%%%%

In order to better understand the influence of the linearized optomechanical-coupling strengths $G_{s1}$ and $G_{s2}$ on the final mean phonon numbers, in Fig.~\ref{Fig8}(a), we plot the final mean phonon numbers $n^{f}_{1}$ and $n^{f}_{2}$ as functions of the scaled decay rate $\kappa/\omega_{1}$ when the linearized optomechanical-coupling strengths take the values of $G_{s2}/\omega_{1}=4G_{s1}/\omega_{1}=0.08$ and $G_{s1}/\omega_{1}=4G_{s2}/\omega_{1}=0.08$. Here we can see that the ground-state cooling of two mechanical modes can be realized in the resolved-sideband regime. In addition, the values of $n^{f}_{1}$ and $n^{f}_{2}$ are approximately exchanged in these two cases, because the parameters of $G_{s1}$ and $G_{s2}$ in these two cases are just antisymmetric. In Fig.~\ref{Fig8}(b)  we plot the final mean phonon numbers $n^{f}_{1}$ and $n^{f}_{2}$ as functions of the ratio $G_{s2}/G_{s1}$ in the case of the linearized optomechanical-coupling strength $G_{s1}/\omega_{1}=0.08$. When $G_{s2}/G_{s1}\leq1$ ($G_{s2}/G_{s1}\geq1$), the final mean phonon numbers of the two mechanical modes increase (decrease) with the increase of the ratio $G_{s2}/G_{s1}$, i.e., the cooling performance of the two mechanical modes is exchanged at the point $G_{s2}/G_{s1}=1$. Due to the dark-mode effect, the ground-state cooling of the two mechanical modes is unfeasible for finite values of the ratio $G_{s2}/G_{s1}$. In this case, the ground-state cooling can be realized by choosing $G_{s2}/G_{s1}<0.4$.

\subsection{The universal conditions for breaking the dark mode \label{darkmode}}

In this section, we analyze the parameter conditions under which the dark mode is formed in the network-coupled optomechanical system. We also study the method for breaking the dark-mode effect. Based on Eqs.~(\ref{linzedlangfull}), we can derive the approximate linearized Hamiltonian. By discarding the noise terms, the linearized optomechanical Hamiltonian under the rotating-wave approximation can be written as
\begin{eqnarray}
\tilde{H}_{\text{RWA}} &=&\Delta_{c}^{\prime}\delta a^{\dagger}\delta a+\Delta_{s}^{\prime\prime}\delta a_{s}^{\dagger}\delta a_{s}+J(\delta a^{\dagger}\delta a_{s}+\delta a_{s}^{\dagger }\delta a)  \notag \\
&&+\sum_{l=1,2}\left[\omega_{l}\delta b_{l}^{\dagger}\delta b_{l}+G_{l}(\delta a\delta b_{l}^{\dagger}+\delta a^{\dagger}\delta b_{l})\right.  \notag \\
&&\left. +G_{sl}(\delta a_{s}\delta b_{l}^{\dagger}+\delta a_{s}^{\dagger}\delta b_{l})\right] +\eta(\delta b_{1}^{\dagger}\delta b_{2}+\delta b_{2}^{\dagger}\delta b_{1}).\notag \\ \label{Hamiltfull}
\end{eqnarray}

By substituting the operators $B_{+}$ ($B^{\dag}_{+}$) and $B_{-}$ ($B^{\dag}_{-}$) into Eq.~(\ref{Hamiltfull}), the Hamiltonian $\tilde{H}_{\text{RWA}}$ becomes [see Fig.~\ref{Fig5}(b)]
\begin{eqnarray}
\tilde{H}_{\text{RWA}}&=&\Delta_{c}^{\prime}\delta a^{\dagger}\delta a+\Delta_{s}^{\prime\prime}\delta a_{s}^{\dagger}\delta a_{s}+J(\delta a^{\dagger}\delta a_{s}+\delta a_{s}^{\dagger}\delta a)+\tilde{\omega}_{+}B_{+}^{\dagger}B_{+}\nonumber\\
&&+\tilde{\omega}_{-}B_{-}^{\dagger }B_{-}+\tilde{\zeta}(B_{+}^{\dagger}B_{-}+B_{-}^{\dagger}B_{+})+G_{+}(\delta aB_{+}^{\dagger}+B_{+}\delta a^{\dagger}) \nonumber\\
&&+\tilde{G}_{s+}(\delta a_{s}B_{+}^{\dagger}+\delta a_{s}^{\dagger}B_{+})+\tilde{G}_{s-}(\delta a_{s}B_{-}^{\dagger}+\delta a_{s}^{\dagger}B_{-}),\label{Hamitfulldig}
\end{eqnarray}
where the resonance frequencies of the two hybrid modes should be replaced by
\begin{subequations}
\label{omegazfnet}
\begin{align}
\tilde{\omega}_{+} =&\frac{\omega_{1}G^{2}_{1}+\omega_{2}G^{2}_{2}+2\eta G_{1}G_{2}}{G^{2}_{+}},\\
\tilde{\omega}_{-} =&\frac{\omega_{1}G^{2}_{2}+\omega_{2}G^{2}_{1}-2\eta G_{1}G_{2}}{G^{2}_{+}},\\ \notag
\end{align}
\end{subequations}
and the three new coupling strengths $\tilde{\zeta}$, $\tilde{G}_{s+}$, and $\tilde{G}_{s-}$ are defined by
\begin{subequations}
\label{jtgszf}
\begin{align}
\tilde{\zeta}&=\frac{(\omega_{1}-\omega_{2})G_{1}G_{2}+\eta(G_{2}^{2}-G_{1}^{2})}{G^{2}_{1}+G^{2}_{2}},\\
\tilde{G}_{s+}&=\frac{G_{s1}G_{1}+G_{s2}G_{2}}{\sqrt{G^{2}_{1}+G^{2}_{2}}},\\
\tilde{G}_{s-}&=\frac{G_{s1}G_{2}-G_{s2}G_{1}}{\sqrt{G^{2}_{1}+G^{2}_{2}}}.\\  \notag
\end{align}
\end{subequations}

%%%%%%%%%%%%%%%%%%%%%
\begin{figure}[tbp]
\center
\includegraphics[width=0.47\textwidth]{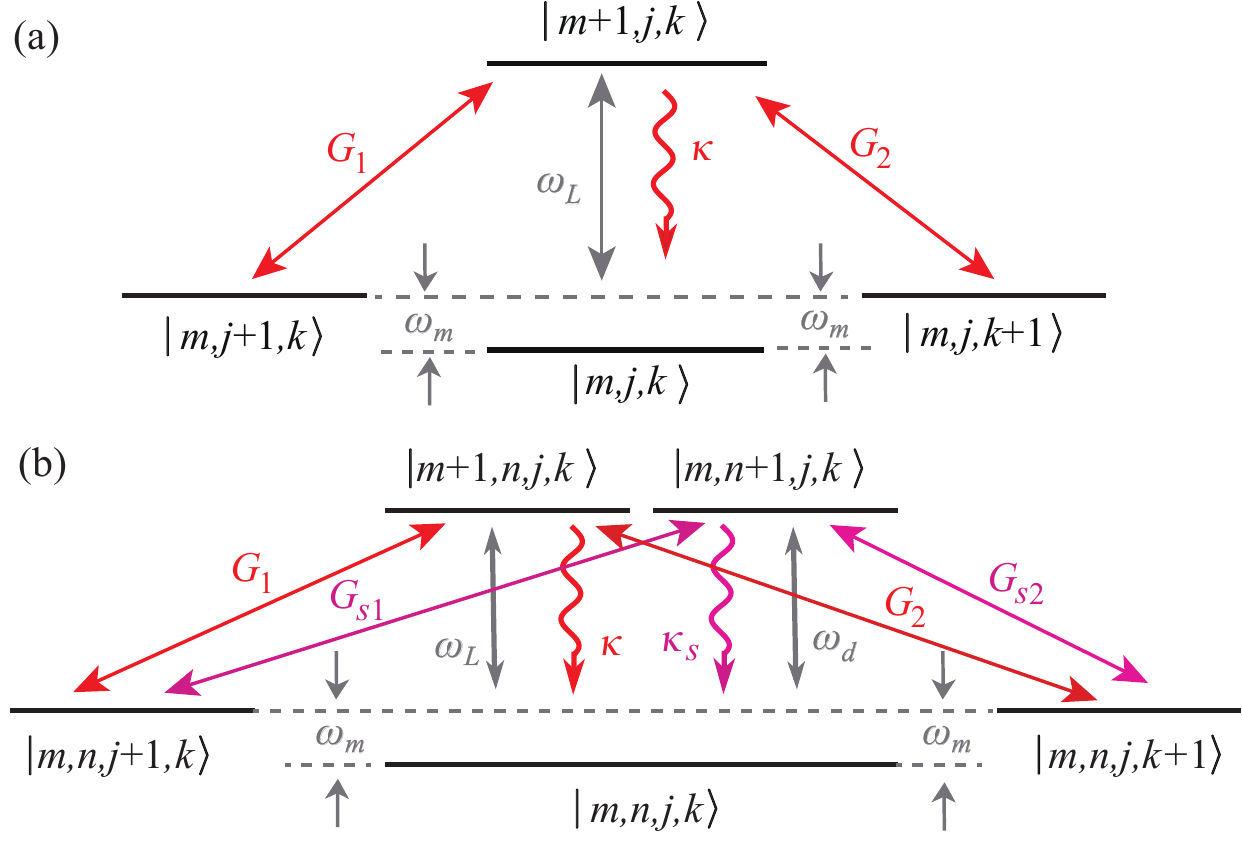}
\caption{Part of the energy levels and transitions of the system in either the (a) absence or (b) presence of the auxiliary cavity mode $a_{s}$. Here, the states $|m,j,k\rangle$ and $|m,n,j,k\rangle$ correspond to the states $|m,j,k\rangle_{a,b_{1},b_{2}}$ and $|m,n,j,k\rangle_{a,a_{s},b_{1},b_{2}}$, respectively. The states $|m\rangle_{a}$, $|n\rangle_{a_{s}}$, $|j\rangle_{b_{1}}$, and $|k\rangle_{b_{2}}$ denote the number states of the cavity mode $a$, auxiliary cavity mode $a_{s}$, mechanical mode $b_{1}$, and mechanical mode $b_{2}$, respectively.}
\label{Fig9}
\end{figure}
%%%%%%%%%%%%%%%%%%%%%%%

From Eqs.~(\ref{Hamitfulldig}) and~(\ref{jtgszf}) we can see that the two hybrid modes $B_{-}$ and $B_{+}$ are decoupled from each other when $\omega_{1}=\omega_{2}$ and $\eta=0$ ($G_{1}=G_{2}$). Moreover, the hybrid mode $B_{-}$ also decouples from the auxiliary cavity mode $a_{s}$ when $\tilde{G}_{s-}=0$. In this case, the hybrid mode $B_{-}$ becomes a dark mode, and the ground-state cooling of the two mechanical modes can not be realized. All in all, the parameter conditions for the appearance of the dark mode are $\tilde{\zeta}=0$ and $\tilde{G}_{s-}=0$, which lead to the conditions:
\begin{subequations}
\label{conditions}
\begin{align}
(\omega_{1}-\omega_{2})G_{1}G_{2}+\eta(G_{2}^{2}-G_{1}^{2})&=0,\\
G_{s1}G_{2}-G_{s2}G_{1}&=0.\\  \notag
\end{align}
\end{subequations}
Therefore, the universal dark-mode-breaking condition is that both $\tilde{\zeta}$ and $\tilde{G}_{s-}$ cannot be zero at the same time [see Fig.~\ref{Fig5}(b)]. In the following we analyze various cases in which the dark mode appears or disappears in the degenerate-resonator case ($\omega_{1}=\omega_{2}$).

(i) In the case of $\eta=0$, i.e., $\tilde{\zeta}=0$, it can be seen from Eq.~(\ref{jtgszf}) that when $G_{s1}G_{2}-G_{s2}G_{1}=0$ ($G_{s1}/G_{s2}=G_{1}/G_{2}$), the hybrid mechanical mode $B_{-}$ (the dark mode) decouples from both the cavity mode $a$ and the auxiliary cavity mode $a_{s}$. In this situation, the excitation energy is stored in the dark mode and cannot be extracted through the cooling channel. In this case, the parameter $J$ has no effect on the breaking of the dark mode effect. This analysis is consistent with the results obtained by numerical calculations in Sec.~\ref{coolingnet}.

(ii) In the case of $\eta\neq 0$ and $G_{1}\neq G_{2}$, i.e., $\tilde{\zeta}\neq0$. we can find that the dark mode can be broken regardless of whether the auxiliary cavity mode appears or not (namely, there is no dark mode).

(iii) In the case of $\eta\neq 0$ and $G_{1}=G_{2}$, i.e., $\tilde{\zeta}=0$, there are two different situations. First, in the case of symmetric coupling ($G_{s1}=G_{s2}$), we can see from Eq.~(\ref{jtgszf}) that $\tilde{G}_{s-}=0$. At this time, the hybrid mechanical mode $B_{-}$ decouples from both the cavity mode $a$ and the auxiliary cavity mode $a_{s}$, i.e., $B_{-}$ becomes a dark mode. However, in the case of asymmetric coupling ($G_{s1}\neq G_{s2}$), i.e., $\tilde{G}_{s-}\neq0$, we can see that the two hybrid mechanical modes $B_{-}$ and $B_{+}$ are coupled with the auxiliary cavity mode $a_{s}$. Even if the hybrid mode $B_{-}$ is decoupled from both the cavity mode $a$ and the hybrid mode $B_{+}$ at the same time, the ground-state cooling of the two mechanical resonators becomes accessible through the cooling channel associated with auxiliary cavity mode $a_{s}$. Obviously, when one of the two coupling strengths $G_{s1}$ and $G_{s2}$ is $0$, the dark-mode effect can naturally be broken. This analysis is consistent with the results we obtained in Sec.~\ref{coolingnet}. Generally speaking, to break the dark mode formed in a three-mode optomechanical system consisting of a cavity mode and two mechanical modes, the easiest way is to introduce an auxiliary cavity mode to couple with one of the two mechanical modes.

\subsection{Analyzing the quantum interference effect in the energy-level transitions of the optomechanical system \label{energy}}

To clarify the physical mechanism behind the dark-mode breaking, in this section we analyze the quantum interference effect in the energy-level transitions of the system. For the optomechanical system, the cavity modes provide the cooling channels to extract the thermal excitations from the mechanical resonators. However, when a single cavity mode is used to cool multiple degenerate mechanical resonators, the phonon modes decaying through the same cooling channel will interfere with each other, similar to the quantum interference effect in electromagnetic induced transparency~\cite{Boller1991,Liu2022}. In this case, some of the mechanical normal modes are decoupled from the cavity mode, then the cooling channel of these decoupled modes (dark modes) is closed and the excitations stored in these mechanical dark modes cannot be extracted. As a result, these dark modes cannot be cooled to their ground states. When multiple cooling channels exist, the phonon dissipation prohibited by one cooling channel can take place via another cooling channel, then the dark-mode effect is broken and the ground-state cooling of multiple mechanical resonators can be realized.

To further understand the quantum interference effect in the energy-level transitions of the optomechanical system, in Figs.~\ref{Fig9}(a) and~\ref{Fig9}(b) we plot the energy levels and transitions of the system including two mechanical modes in either the absence or presence of the auxiliary cavity mode $a_{s}$. For simplicity, we ignore the subscripts of the basis vectors, i.e., we denote $|m,j,k\rangle_{a,b_{1},b_{2}}$ ($|m,n,j,k\rangle_{a,a_{s},b_{1},b_{2}}$) by $|m,j,k\rangle$ ($|m,n,j,k\rangle$). As shown in Fig.~\ref{Fig9}(a), under the red-sideband-resonance condition $\Delta_{c}=\omega_{m}$, the transitions $|m,j+1,k\rangle\leftrightarrow|m+1,j,k\rangle$ and $|m,j,k+1\rangle\leftrightarrow|m+1,j,k\rangle$ are resonant and the transition $|m+1,j,k\rangle\rightarrow|m,j,k\rangle$ can further occur through the cavity-field decay $\kappa$. When the mechanical modes $b_{1}$ and $b_{2}$ are degenerate, the phonon modes decaying through the same cooling channel (cavity mode $a$) will interfere destructively with each other; then the ground-state cooling of the two mechanical modes cannot be realized. To break the dark-mode effect in this system, a natural and simple method is to introduce another cooling channel to the mechanical resonators. As shown in Fig.~\ref{Fig9}(b), under the red-sideband-resonance condition $\Delta_{c}=\omega_{m}$, except for the transitions $|m,n,j+1,k\rangle\leftrightarrow|m+1,n,j,k\rangle$ and $|m,n,j,k+1\rangle\leftrightarrow|m+1,n,j,k\rangle$ associated with the cavity mode $a$, the transitions $|m,n,j+1,k\rangle\leftrightarrow|m,n+1,j,k\rangle$ and $|m,n,j,k+1\rangle\leftrightarrow|m,n+1,j,k\rangle$ associated with the auxiliary cavity mode $a_{s}$ can also occur. In this way, phonon dissipation prohibited by the cooling channel $a$ can be decayed via a new cooling channel $a_{s}$. Therefore, by introducing the auxiliary cavity mode $a_{s}$, the dark-mode effect can be broken and the ground-state cooling of two mechanical resonators can be realized.

We point out that the main innovation in this work is the breaking of the dark-mode effect via the auxiliary cavity and that the underlying physical mechanism of this scheme is the sideband cooling. Therefore, the system should work in the resolved-sideband regime and a resonant red-sideband driving is needed. In addition, though the dark mode appears theoretically in the degenerate mechanical-resonator case, our scheme works well even for near-degenerate resonators within a detuning window with the width of the cavity-field decay rate. Physically, the resonators with these detuned frequencies cannot be distinguished via the cavity emission spectrum.

\section{Ground-state cooling and dark-mode breaking in a multiple-mechanical-mode optomechanical system\label{Coolmulti}}

%%%%%%%%%%%%%%%%%%%%%%%%%%%%%%
\begin{figure}[tbp]
\center
\includegraphics[width=0.47 \textwidth]{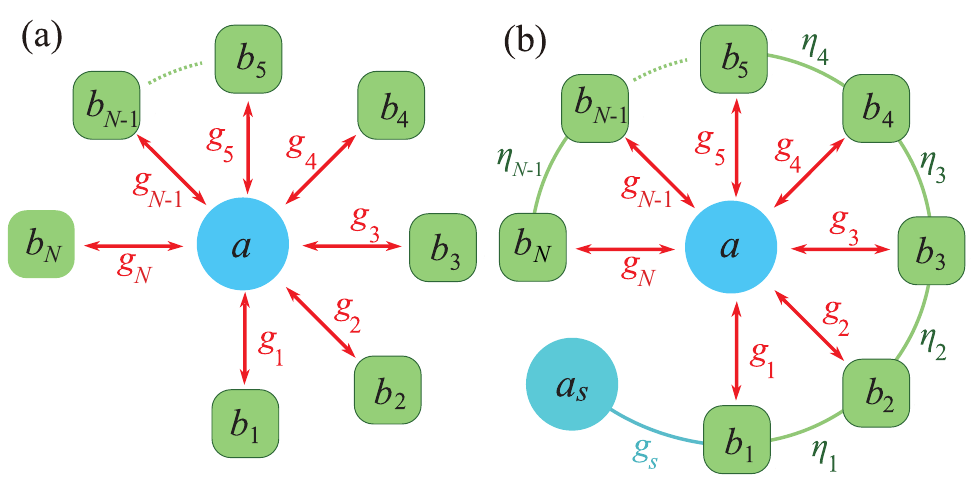}
\caption{(a) $N$-mechanical-mode optomechanical system consisting of an intermediate cavity mode optomechanically coupled to $N$ mechanical modes. (b) Both the phonon-hopping interactions between two neighboring mechanical modes and the optomechanical interaction between the auxiliary cavity mode and the mechanical mode $b_{1}$ are introduced into the $N$-mechanical-mode optomechanical system.}
\label{Fig10}
\end{figure}
%%%%%%%%%%%%%%%%%%%%%%%%%%%%%%

In this section, we generalize the auxiliary-cavity-mode method to realize the simultaneous ground-state cooling of $N$ mechanical modes in a multiple-mechanical-mode optomechanical system, in which an intermediate cavity mode is coupled to $N$ ($N\geq 3$) mechanical modes [see Fig.~\ref{Fig10}(a)]. Concretely, we introduce an auxiliary cavity mode optomechanically coupled to the first mechanical mode. We also introduce the phonon-hopping coupling between all the neighboring two mechanical modes [see Fig.~\ref{Fig10}(b)]. Moreover, we analyze the parameter conditions for forming and breaking the dark modes.

\subsection{Simultaneous ground-state cooling of $N$ mechanical modes}

We consider a multiple-mechanical-mode optomechanical system, which is composed of an intermediate cavity mode, an auxiliary cavity mode, and $N$ ($N\geq 3$) mechanical modes [see Fig.~\ref{Fig10}(b)]. In a rotating frame defined by the transformation operator $\exp[-i(\omega_{L}a^{\dagger}a+\omega_{d}a_{s}^{\dagger}a_{s})t]$, the Hamiltonian of this system is written as
\begin{eqnarray}
H_{I}&=&\Delta_{c}a^{\dagger}a+\sum_{l=1}^{N}[\omega_{l}b_{l}^{\dagger}b_{l}+g_{l}a^{\dagger }a(b_{l}^{\dagger}+b_{l})]\nonumber\\
&&+\sum_{l=1}^{N-1}\eta_{l}(b_{l}^{\dagger}b_{l+1}+b_{l+1}^{\dagger}b_{l})+(\Omega a^{\dagger}+\mathrm{H.c.})\nonumber\\
&&+\Delta_{s}a_{s}^{\dagger}a_{s}+g_{s}a_{s}^{\dagger}a_{s}(b_{1}^{\dagger}+b_{1})+(\Omega_{s}a_{s}^{\dagger}+\mathrm{H.c.}),\label{Hamitmulti}
\end{eqnarray}
where the operators and variables for the cavity modes have been defined before, and $b_{l}$ ($b^{\dagger}_{l}$) are the annihilation (creation) operators of the $l$th mechanical mode. The parameter $g_{l}$ ($g_{s}$) denotes the single-photon optomechanical-coupling strength between the intermediate (auxiliary) cavity mode and the $l$th mechanical mode (mechanical mode $b_{1}$). The cavity mode $a$ $(a_{s})$ is strongly driven by the laser field with the driving frequency $\omega_{L}$ ($\omega_{d}$) and the driving amplitude $\Omega$ ($\Omega_{s}$).

Similar to the two-mechanical-mode case, the two cavity fields are strongly driven by two laser fields, then the dynamics of this system can be treated by the linearization method. Based on the Hamiltonian~(\ref{Hamitmulti}), the linearized Langevin equations for the quantum fluctuations are given by
\begin{eqnarray}
\delta\dot{a} &=&-(\kappa+i\tilde{\Delta}_{c})\delta a-i[g_{1}\alpha(\delta b_{1}+\delta b_{1}^{\dagger})
+g_{2}\alpha(\delta b_{2}+\delta b_{2}^{\dagger})+\ldots\nonumber \\
&&+ig_{N}\alpha (\delta b_{N}+\delta b_{N}^{\dagger})] +\sqrt{2\kappa}a_{\text{in}}, \nonumber \\
\delta\dot{a}_{s} &=&-(\kappa_{s}+i\Delta_{s}^{\prime})\delta a_{s}-ig_{s}\alpha_{s}(\delta b_{1}+\delta b_{1}^{\dagger})+\sqrt{2\kappa_{s}}a_{s,\text{in}}, \nonumber \\
\delta\dot{b}_{1} &=&-(\gamma_{1}+i\omega_{1})\delta b_{1}-ig_{1}\alpha^{\ast}\delta a-ig_{1}\alpha\delta a^{\dagger}-ig_{s}\alpha_{s}^{\ast}\delta a_{s}\nonumber \\
&&-ig_{s}\alpha_{s}\delta a_{s}^{\dag}-i\eta_{1}\delta b_{2}+\sqrt{2\gamma_{1}}b_{1,\text{in}}, \nonumber \\
\delta\dot{b}_{2} &=&-(\gamma_{2}+i\omega_{2})\delta b_{2}-ig_{2}\alpha^{\ast}\delta a-ig_{2}\alpha\delta a^{\dagger}-i\eta_{1}\delta b_{1}\nonumber \\
&&-i\eta_{2}\delta b_{3}+\sqrt{2\gamma_{2}}b_{2,\text{in}}, \nonumber \\
\delta\dot{b}_{3} &=&-(\gamma_{3}+i\omega_{3})\delta b_{3}-ig_{3}\alpha^{\ast}\delta a-ig_{3}\alpha\delta a^{\dagger}-i\eta_{2}\delta b_{2} \nonumber \\
&&-i\eta_{3}\delta b_{4}+\sqrt{2\gamma_{3}}b_{3,\text{in}}, \nonumber \\
&&\vdots  \nonumber \\
\delta \dot{b}_{N-1} &=&-(\gamma_{N-1}+i\omega_{N-1})\delta b_{N-1}-ig_{N-1}\alpha^{\ast}\delta a-ig_{N-1}\alpha\delta a^{\dagger}\nonumber \\
&&-i\eta_{N-2}\delta b_{N-2}-i\eta_{N-1}\delta b_{N}+\sqrt{2\gamma_{N-1}}b_{N-1,\text{in}}, \nonumber \\
\delta\dot{b}_{N} &=&-(\gamma_{N}+i\omega_{N})\delta b_{N}-ig_{N}\alpha^{\ast}\delta a-ig_{N}\alpha\delta a^{\dagger}\nonumber \\
&&-i\eta_{N-1}\delta b_{N-1}+\sqrt{2\gamma_{N}}b_{N,\text{in}},
\label{linearmulti}
\end{eqnarray}
where $\tilde{\Delta}_{c}=\Delta_{c}+\sum_{l=1}^{N}g_{l}(\beta_{l}+\beta_{l}^{\ast})$ [$\Delta_{s}^{\prime}=\Delta_{s}+g_{s}(\beta_{1}+\beta_{1}^{\ast})$] is the effective driving detuning of the cavity mode $a$ ($a_{s}$) and $G_{l}=g_{l}|\alpha|$ ($G_{s}=g_{s}|\alpha_{s}|$) is the linearized optomechanical-coupling strength between the cavity mode $a$ ($a_{s}$) and the $l$th mechanical mode (mechanical mode $b_{1}$).

Below we study the ground-state cooling of $N$ mechanical modes. The cooling performance of mechanical modes can be verified by calculating the final mean phonon numbers. Therefore, we rewrite the linearized Langevin equations~(\ref{linearmulti}) as
\begin{eqnarray}
\mathbf{\dot{\tilde{u}}}(t)=\mathbf{\tilde{A}\tilde{u}}(t)+\mathbf{\tilde{N}}(t),\label{MatrixLangevinN}
\end{eqnarray}
where we introduce the vector of the quantum fluctuations
$\mathbf{\tilde{u}}(t)=[\delta a,\delta a_{s},\delta b_{1},\ldots,\delta b_{N-1},\delta b_{N},\delta a^{\dagger},\delta a_{s}^{\dagger},\delta b_{1}^{\dagger},\ldots,\delta b^{\dagger}_{N-1},\delta b^{\dagger}_{N}]^{T}$ and the vector of the quantum noise $\mathbf{\tilde{N}}(t)=\sqrt{2}[\sqrt{\kappa}a_{\text{in}},\sqrt{\kappa_{s}}a_{s,\text{in}},\sqrt{\gamma_{1}}b_{1,\text{in}},\ldots,\sqrt{\gamma _{N-1}}b_{N-1,\text{in}},\sqrt{\gamma_{N}}b_{N,\text{in}}, \\ \sqrt{\kappa}a_{\text{in}}^{\dagger},\sqrt{\kappa_{s}}a_{s,\text{in}}^{\dag},\sqrt{\gamma_{1}}b_{1,\text{in}}^{\dag},\ldots,\sqrt{\gamma_{N-1}}
b_{N-1,\text{in}}^{\dag},\sqrt{\gamma_{N}}b_{N,\text{in}}^{\dag}]^{T}$. The corresponding coefficient matrix
$\mathbf{\tilde{A}}=\left(
\begin{array}{cc}
-\mathbf{\tilde{E}} & \mathbf{\tilde{F}} \\
\mathbf{\tilde{F}}^{\ast} & -\mathbf{\tilde{E}}^{\ast}
\end{array}
\right)$,
where
\small
%\footnotesize
\begin{eqnarray}
\mathbf{\tilde{E}}
&=&\left(
\begin{array}{cccccc}
\kappa +i\tilde{\Delta}_{c}  & 0 & iG_{1} & \cdots & iG_{N-1} & iG_{N} \\
0 & \kappa_{s}+i\Delta_{s}^{\prime}  & iG_{s} & \cdots & 0 & 0 \\
iG_{1}^{\ast} & iG_{s}^{\ast} & \gamma_{1}+i\omega_{1} & \cdots & 0 & 0 \\
\vdots & \vdots & \vdots & \ddots & \vdots & \vdots \\
iG_{N-1}^{\ast} & 0 & 0 & \cdots & \gamma_{N-1}+i\omega_{N-1} & i\eta_{N-1} \\
iG_{N}^{\ast} & 0 & 0 & \cdots & i\eta_{N-1} & \gamma_{N}+i\omega_{N}
\end{array}
\right)\nonumber \\
\end{eqnarray}
\normalsize
and
\begin{equation}
\mathbf{\tilde{F}}=\left(
\begin{array}{cccccc}
0 & 0 & -iG_{1} & \cdots & -iG_{N-1} & -iG_{N} \\
0 & 0 & -iG_{s} & \cdots & 0 & 0 \\
-iG_{1} & -iG_{s} & 0 & \cdots & 0 & 0 \\
\vdots & \vdots & \vdots & \ddots  & \vdots & \vdots \\
-iG_{N-1} & 0 & 0 & \cdots & 0 & 0 \\
-iG_{N} & 0 & 0 & \cdots & 0 & 0
\end{array}
\right).
\end{equation}
The formal solution of Eq.~(\ref{MatrixLangevinN}) is given by
\begin{equation}
\mathbf{\tilde{u}}(t) =\mathbf{\tilde{M}}(t) \mathbf{\tilde{u}}(0)+\int_{0}^{t}\mathbf{\tilde{M}}(t-s) \mathbf{\tilde{N}}(s)ds,\label{FormalsoluN}
\end{equation}
where the matrix $\mathbf{\tilde{M}}(t)=\text{exp}(\mathbf{\tilde{A}}t)$. Note that our simulations should satisfy the stability conditions, which can
be obtain by analyzing the Routh-Hurwitz criterion~\cite{Gradstein2014}.

Based on Eq.~(\ref{FormalsoluN}), we can obtain the final mean phonon numbers by solving the Lyapunov equation. For this reason, we introduce the covariance matrix $\mathbf{\tilde{V}}$ of the system by defining the matrix elements as
\begin{equation}
\mathbf{\tilde{V}}_{ij}=\frac{1}{2}[\langle\mathbf{\tilde{u}}_{i}(\infty)\mathbf{\tilde{u}}_{j}(\infty)\rangle+\langle\mathbf{\tilde{u}}_{j}(\infty)\mathbf{\tilde{u}}_{i}(\infty)\rangle]\label{CovarianceN}.
\end{equation}
Under the stability conditions, the covariance matrix $\mathbf{\tilde{V}}$ is determined by the Lyapunov equation
\begin{equation}
\mathbf{\tilde{A}}\mathbf{\tilde{V}}+\mathbf{\tilde{V}}\mathbf{\tilde{A}}^{T}=-\mathbf{\tilde{Q}},
\end{equation}
where
\begin{equation}
\mathbf{\tilde{Q}}=\frac{1}{2}(\mathbf{\tilde{C}}+\mathbf{\tilde{C}}^{T}),
\end{equation}
with the correlation matrix $\mathbf{\tilde{C}}$ related to the noise operators. In the Markovian-dissipation case, the correlation matrix $\mathbf{\tilde{C}}$ can be obtained as
$\mathbf{\tilde{C}}=2\left(
\begin{array}{cc}
0 & \mathbf{P} \\
\mathbf{R} & 0
\end{array}
\right)$,
where
\begin{equation}
\mathbf{P}=\left(
\begin{array}{cccccc}
\kappa & 0 & 0 & \cdots & 0 & 0 \\
0 & \kappa_{s} & 0 & \cdots & 0 & 0 \\
0 & 0 & \gamma_{1}\left(\bar{n}_{1}+1\right) & \cdots & 0 & 0 \\
\vdots & \vdots & \vdots & \ddots & \vdots & \vdots \\
0 & 0 & 0 & \cdots & \gamma_{N-1}\left(\bar{n}_{N-1}+1\right) & 0 \\
0 & 0 & 0 & \cdots & 0 & \gamma_{N}\left(\bar{n}_{N}+1\right)  \\
\end{array}
\right)
\end{equation}
and
\begin{equation}
\mathbf{R}=\left(
\begin{array}{cccccc}
0 & 0 & 0 & \cdots & 0 & 0 \\
0 & 0 & 0 & \cdots & 0 & 0 \\
0 & 0 & \gamma_{1}\bar{n}_{1} & \cdots & 0 & 0 \\
\vdots & \vdots & \vdots & \ddots  & \vdots & \vdots \\
0 & 0 & 0 & \cdots & \gamma_{N-1}\bar{n}_{N-1} & 0 \\
0 & 0 & 0 & \cdots & 0 & \gamma_{N}\bar{n}_{N}
\end{array}
\right).
\end{equation}

According to the covariance matrix $\mathbf{\tilde{V}}$ defined in Eq.~(\ref{CovarianceN}), we can derive the final mean phonon numbers of the $l$th mechanical mode as
\begin{eqnarray}
\label{finalexactNmr}
\langle \delta b_{l}^{\dagger}\delta b_{l}\rangle=\mathbf{\tilde{V}}_{N+l+4,l+2}-\frac{1}{2},
\end{eqnarray}
where $\mathbf{\tilde{V}}_{N+l+4,l+2}$ is the matrix element of the covariance matrix $\mathbf{\tilde{V}}$.

%%%%%%%%%%%%%%%%%%%%%%%%%%%%%%
\begin{figure}[tbp]
\centering
\includegraphics[width=0.47 \textwidth]{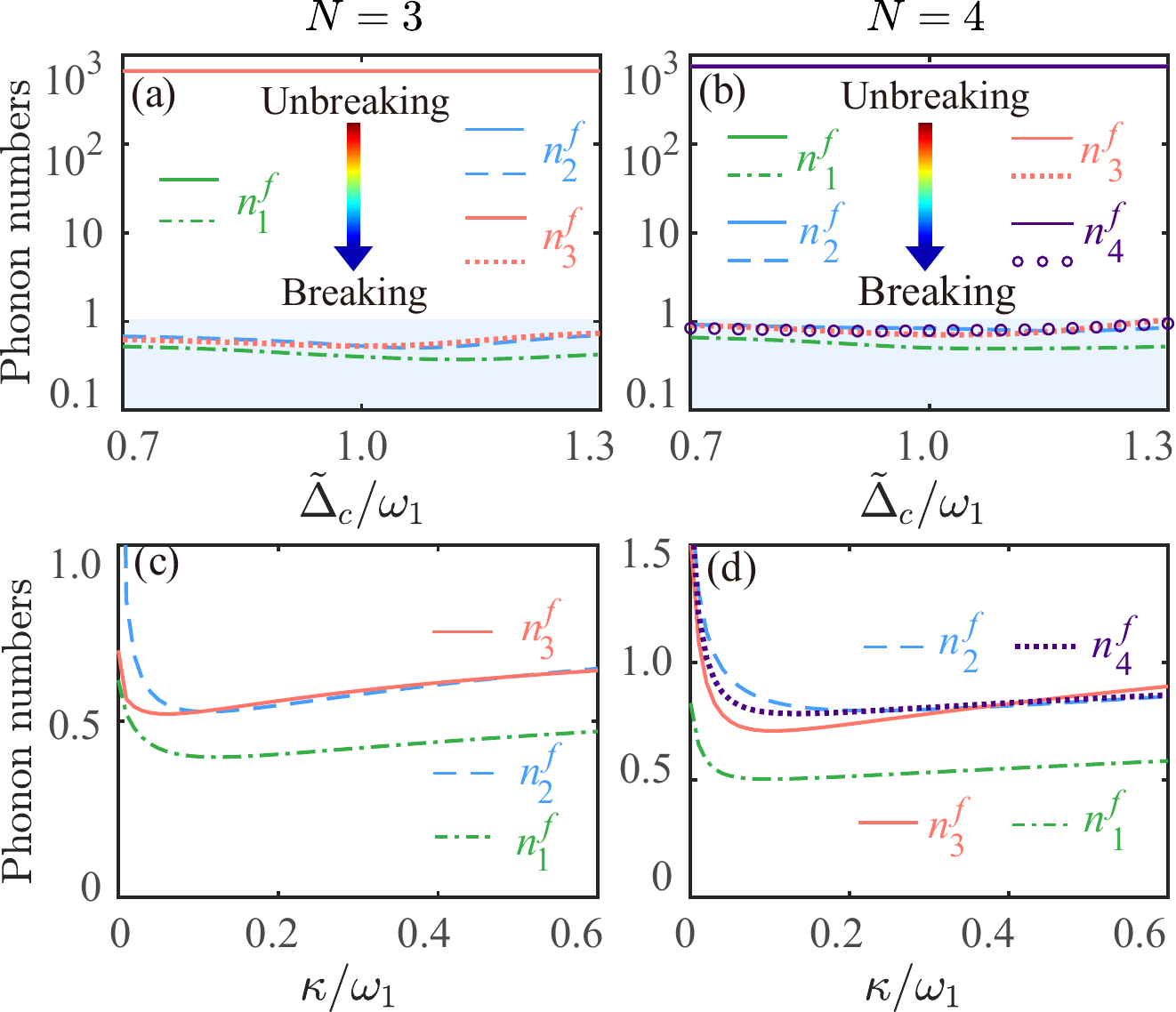}
\caption{(a) and (b) Final mean phonon numbers  $n^{f}_{l}$ as functions of the scaled driving detuning $\tilde{\Delta}_{c}/\omega_{1}$ in the dark-mode-unbreaking ($G_{s}/\omega_{1}=0$ and $\eta_{l}/\omega_{1}=0$) or -breaking ($G_{s}/\omega_{1}=0.1$ and $\eta_{l}/\omega_{1}=0.06$) cases  for (a) $N = 3$ and (b) $N = 4$. (c) and  (d) Final mean phonon numbers  $n^{f}_{l}$ as functions of the scaled cavity-field decay rate $\kappa/\omega_{1}$ for (a) $N = 3$ and (b) $N = 4$. The other parameters are $\omega_{l}/\omega_{1}=1$, $\Delta_{s}^{\prime}/\omega_{1}=1$, $\kappa_{s}/\omega_{1}=0.1$, $\eta_{l}/\omega_{1}=0.06$, $\gamma_{l}/\omega_{1}=10^{-5}$, $G_{l}/\omega_{1}=0.05$, $G_{s}/\omega_{1}=0.1$, $\bar{n}_{l}=1000$, and (a) and (b) $\kappa/\omega_{1}=0.1$ and (c) and (d) $\tilde{\Delta}_{c}/\omega_{1}=1$.}
\label{Fig11}
\end{figure}
%%%%%%%%%%%%%%%%%%%%%%%%%%%%%%

Below we study the cooling performance of the mechanical modes when there are three or four mechanical modes in the optomechanical system. For simplicity, we consider that the resonance frequencies of all mechanical modes are equal ($\omega_{l}=\omega_{m}$ for $l=1,\ldots,N$). Similarly, the phonon-hopping coupling strengths [$\eta_{l}=\eta$ for $l=1,\ldots,(N-1)$] and optomechanical-coupling strengths ($G_{l}=G$ for $l=1,\ldots,N$) are also the same.

As shown in Figs.~\ref{Fig11}(a) and ~\ref{Fig11}(b), the final mean phonon numbers $n^{f}_{j}$ are plotted as functions of the scaled driving detuning $\tilde{\Delta}_{c}/\omega_{m}$ in both the dark-mode-breaking case ($G_{s}/\omega_{1}=0.1$  and $\eta_{l}/\omega_{m}=0.06$) and the dark-mode-unbreaking case ($G_{s}=\eta_{l}=0$). We can see that the ground-state cooling cannot be realized for these mechanical modes when the auxiliary cavity mode and phonon-hopping interactions are absent ($G_{s}=\eta_{l}=0$) [see the upper curves in Figs.~\ref{Fig11}(a) and~\ref{Fig11}(b)]. This is because the thermal excitations stored in the dark modes cannot be extracted through the cooling channel related to cavity mode $a$.

When the auxiliary cavity mode and the phonon-hopping interactions are introduced, a new cooling channel is formed and the dark modes are broken. Then the ground-state cooling of multiple mechanical modes can be realized, as shown in Figs.~\ref{Fig11}(a) and~\ref{Fig11}(b). Moreover, in Figs.~\ref{Fig11}(c) and ~\ref{Fig11}(d)  we plot the final mean phonon numbers $n^{f}_{l}$ as functions of the cavity-field decay rate $\kappa/\omega_{1}$. We find that the cooling performance of the first mechanical mode is the best, because it is directly connected to the auxiliary cavity mode. The cooling performance of other mechanical modes is almost the same, because all the other mechanical modes have similar coupling connections with the cooling baths.

\subsection{Parameter conditions for breaking the dark modes}

To study the parameter conditions for breaking the dark modes, we can derive the approximate linearized Hamiltonian, which governs the dynamics of the system. To implement the cooling scheme, the system should work in the red-sideband resonance regime, in which the rotating-wave approximation can be safely performed. By discarding the noise terms, the linearized optomechanical Hamiltonian is given by
\begin{eqnarray}
H_{I}&=&\tilde{\Delta}_{c}\delta a^{\dagger}\delta a+\sum_{l=1}^{N}\omega_{l}\delta b_{l}^{\dagger}\delta b_{l}+\sum_{l=1}^{N}G_{l}(\delta a^{\dagger}\delta b_{l}+\delta b_{l}^{\dagger}\delta a) \nonumber\\
&&+\Delta_{s}^{\prime}\delta a_{s}^{\dagger}\delta a_{s}+\sum_{l=1}^{N-1}[\eta_{l}(\delta b_{l}\delta b_{l+1}^{\dagger}+\delta b_{l+1}\delta b_{l}^{\dagger})] \nonumber\\
&&+G_{s}(\delta a_{s}^{\dagger}\delta b_{1}+\delta b_{1}^{\dagger}\delta a_{s}).\label{HamitRWAmuti}
\end{eqnarray}
To clearly see the dark-mode effect in the multiple-mechanical-mode optomechanical system, we first consider the situation where the
auxiliary cavity and the phonon-hopping interactions between two neighboring mechanical modes are absent, i.e., $\Delta_{s}^{\prime}=0$, $\eta_{l}=0$, and $G_{s}=0$ [see Fig.~\ref{Fig10}(a)]. Then the Hamiltonian~(\ref{HamitRWAmuti}) becomes
\begin{equation}
H_{I}^{\prime}=\tilde{\Delta}_{c}\delta a^{\dagger}\delta a+\sum_{l=1}^{N}\omega_{l}\delta b_{l}^{\dagger}\delta b_{l}+\sum_{l=1}^{N}G_{l}(\delta a^{\dagger}\delta b_{l}+\delta b_{l}^{\dagger}\delta a).
\end{equation}
For simplicity, we consider that all the mechanical modes have the same resonance frequencies ($\omega_{l}=\omega_{m}$), and the optomechanical-coupling strengths between the intermediate cavity mode and all mechanical modes are also the same ($G_{l}=G$). In this case, there are a bright mode $B_{+}=\sum_{l=1}^{N}\delta b_{l}/\sqrt{N}$ and ($N-1$) dark modes decoupled from the intermediate cavity mode. Therefore, the thermal excitations stored in the dark modes cannot be extracted though the cooling channel of the cavity mode, and then the ground-state cooling of these mechanical modes cannot be realized.

To break the dark mode and achieve the ground-state cooling of $N$ mechanical modes, we introduce an auxiliary cavity mode optomechanically coupled to the mechanical mode $b_{1}$ and phonon-hopping interactions between the neighboring mechanical modes, as shown in Fig.~\ref{Fig10}(b). For convenience, we consider the case where all the phonon-hopping coupling strengths are the same ($\eta_{l}=\eta$). Thus, the Hamiltonian associated with these coupled mechanical modes can be diagonalized as
\begin{eqnarray}
H_{\text{mph}} &=&\omega_{m}\sum_{l=1}^{N}\delta b_{l}^{\dagger}\delta b_{l}+\eta\sum_{l=1}^{N-1}(\delta b_{l}\delta b_{l+1}^{\dagger}+\delta b_{l+1}\delta b_{l}^{\dagger}) \nonumber\\
&&=\sum_{k=1}^{N}\Omega_{k}B_{k}^{\dagger}B_{k},
\end{eqnarray}
where $\Omega_{k}$ is the resonance frequency of the $k$th hybrid mechanical mode $B_{k}$ and  is defined by
\begin{eqnarray}
\Omega_{k}=\omega_{m}+2\eta\cos \left(\frac{k\pi}{N+1}\right),\hspace{0.5 cm}k=1,2,3,\ldots,N.
\end{eqnarray}
Meanwhile, the relationship between the hybrid mode $B_{k}$ and the mechanical mode $\delta b_{l}$ can be expressed as
\begin{equation}
\delta b_{l}=\frac{1}{D}\sum_{k=1}^{N}\sin \left(\frac{lk\pi}{N+1}\right)B_{k},
\end{equation}
with $D=\sqrt{(N+1)/2}$. When the auxiliary cavity mode is absent, by substituting the hybrid mode $B_{k}$ into the Hamiltonian~(\ref{HamitRWAmuti}), we can obtain
\begin{eqnarray}
H_{I}=&\tilde{\Delta}_{c}\delta a^{\dagger}\delta a+\sum_{k=1}^{N}\Omega_{k}B_{k}^{\dagger}B_{k}+H_{\text{oi}},
\end{eqnarray}
where the optomechanical-interaction Hamiltonian $H_{\text{oi}}$ is given by
\begin{equation}
H_{\text{oi}}=\sum_{k=1}^{N}\left[\frac{G}{D}\sum_{l=1}^{N}\sin\left(\frac{lk\pi}{N+1}\right)\delta aB_{k}^{\dagger}+\mathrm{H.c.}\right].\label{Hamitoi}
\end{equation}
From Eq.~(\ref{Hamitoi}) we can see that the coupling strength between the cavity mode $a$ and the hybrid mode $B_{k}$ is determined by the coefficient $\frac{G}{D}\left\{\sum_{l=2}^{N}\sin\left[lk\pi/(N+1)\right]\right\}$. Hence, we next analyze the dependence of this coefficient on the variables $k$ and $N$.

First, we consider the situation of $N=2$. In this case, the system is reduced to the two-mechanical-mode optomechanical system, which has been analyzed in detail in previous sections. When $N=2$, the optomechanical interaction becomes
\begin{eqnarray}
H_{\text{oi}}&=&\sqrt{2}G \delta a B_{1}^{\dagger}+\sqrt{2}G^{\ast}B_{1} \delta a^{\dagger}.
\end{eqnarray}
It is obvious that the hybrid mechanical mode $B_{2}$ becomes a dark mode, which decouples from both the cavity mode $a$ and the hybrid mechanical mode $B_{1}$, so the ground-state cooling of the two mechanical modes becomes inaccessible.

In the case of $N\geq3$, the coupling coefficient between the cavity mode $a$ and the $k$th hybrid mechanical mode $B_{k}$ defined in Eq.~(\ref{Hamitoi}) is given by $\frac{G}{D}\sum_{l=1}^{N}\sin \left(\frac{lk\pi }{N+1}\right)$. Since the forms of the coupling coefficients are different when $N$ is either an odd number or an even number, below we will analyze two cases corresponding to odd and even $N$, respectively.
(i) When $N$ is an odd number, the form of the coupling coefficient depends on $k$. If $k$ is also an odd number, we get
$\frac{G}{D}\sum_{l=1}^{N}\sin \left(\frac{lk\pi}{N+1}\right)\neq0$. If $k$ is even, we obtain $\frac{G}{D}\sum_{l=1}^{N}\sin \left(\frac{lk\pi }{N+1}\right)=0$.
(ii) For an even $N$, when $k$ is an odd number, we obtain $\frac{G}{D}\sum_{l=1}^{N}\sin \left(\frac{lk\pi}{N+1}\right)\neq0$.
When $k$ is also an even number, we get $\frac{G}{D}\sum_{l=1}^{N}\sin \left(\frac{lk\pi }{N+1}\right)=0$.

Based on the above discussion, we can find that for an odd $k$, the coupling strength between the intermediate cavity mode $a$ and the $k$th hybrid mechanical mode $B_{k}$ is nonzero. However, when $k$ is an even number, the intermediate cavity mode $a$ is decoupled from the $k$th hybrid mechanical mode $B_{k}$. In this situation, all the even hybrid mechanical modes are decoupled from the intermediate cavity mode. Thus, the ground-state cooling of multiple mechanical modes cannot be realized under the influence of the dark-mode effect. Nevertheless, we can introduce an auxiliary cavity mode $a_{s}$ to break the dark-mode effect, which is coupled to the mechanical mode $b_{1}$ via the radiation-pressure interaction. By substituting the hybrid mechanical modes $B_{k}$ into the optomechanical Hamiltonian $H_{\text{som}}=G_{s}(\delta a_{s}^{\dagger}\delta b_{1}+\delta b_{1}^{\dagger}\delta a_{s})$, we can get
\begin{equation}
H_{\text{som}}=G_{s}\sum_{k=1}^{N}\left[\sin \left(\frac{k\pi}{N+1}\right)\delta a_{s}^{\dagger}B_{k}+\mathrm{H.c.}\right].\label{NorHamitsom}
\end{equation}
It can be seen from Eq.~(\ref{NorHamitsom}) that all the hybrid mechanical modes $B_{k}$ are coupled with the auxiliary cavity mode $a_{s}$. Even when the even hybrid mechanical modes are decoupled from the intermediate cavity mode $a$, the ground-state cooling of the $N$ mechanical modes can still be achieved through the cooling channel associated with the auxiliary cavity mode $a_{s}$.

\section{Physical mechanism for breaking the dark-state effect in the ``$N$"-type four-level atomic system\label{mechanism}}

%%%%%%%%%%%%%%%%%%%%%
\begin{figure}[tbp]
\center
\includegraphics[width=0.48\textwidth]{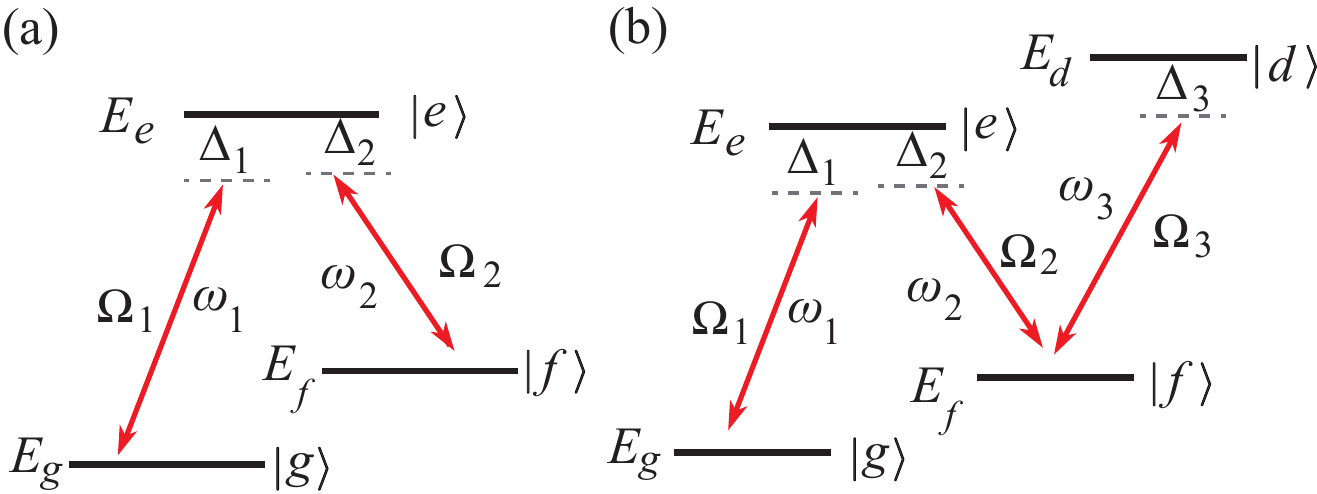}
\caption{(a) Schematic of the three-level system, which consists of three states $\vert e\rangle$, $\vert f\rangle$, and $\vert g\rangle$, with the corresponding energies $E_{e}$, $E_{f}$, and $E_{g}$. The coupling strengthes of the transition processes $|g\rangle \leftrightarrow |e\rangle $ and $|f\rangle \leftrightarrow |e\rangle $ are, respectively, denoted by $\Omega_{1}$ and $\Omega_{2}$ with the corresponding detunings $\Delta_{1}$ and $\Delta_{2}$. (b) Schematic of the $N$-type four-level system. In addition to the couplings and states introduced in (a), we introduce an auxiliary state $|d\rangle$ coupled to the lower state $\vert f\rangle$ with the coupling strength $\Omega_{3}$ and the detuning $\Delta_{3}$.}
\label{Fig12}
\end{figure}
%%%%%%%%%%%%%%%%%%%%%%%
To further investigate the generality of the physical mechanism for breaking the dark-mode effect, in this section we consider the dark-state effect in an atomic-level system. Concretely, we demonstrate the dark-state effect in a $\Lambda$-type three-level system and show how to break this dark-state effect by introducing an auxiliary state coupled to one of the two lower states, namely, forming an $N$-type four-level system (Fig.~\ref{Fig12}). For the three-level system, the Hamiltonian reads
\begin{eqnarray}
H_{\text{TLS}}(t)&=&E_{e}\vert e\rangle \langle e\vert +E_{f}\vert f\rangle \langle f\vert+E_{g}\vert g\rangle\langle g\vert \nonumber\\
&&+(\Omega_{1}\vert e\rangle\langle g\vert e^{-i\omega_{1}t}+\Omega_{2}\vert e\rangle \langle f\vert e^{-i\omega
_{2}t}+\mathrm{H.c.}),\label{Hamitthree}
\end{eqnarray}
where $E_{e}$, $E_{f}$, and $E_{g}$ denote the energies of energy levels $|e\rangle $, $|f\rangle $, and $|g\rangle $, respectively. For convenience, hereafter we assume that the energy of the ground state is $0$ ($E_{g}=0$). The two atomic transitions $|g\rangle \leftrightarrow |e\rangle$  and $|f\rangle \leftrightarrow |e\rangle $ are, respectively, coupled to the two monochromatic fields with the frequencies $\omega_{1}$ and $\omega_{2}$ and the transition amplitudes $\Omega_{1}$ and $\Omega_{2}$. In a rotating frame with respect to $\omega_{1}\vert e\rangle \langle e\vert+E_{f}\vert f\rangle \langle f\vert +E_{g}\vert g\rangle \langle g\vert$, the Hamiltonian in Eq.~(\ref{Hamitthree}) becomes
\begin{eqnarray}
V_{I}=\Delta_{1} \vert e\rangle \langle e\vert +\Omega_{1}(\vert e\rangle \langle g\vert+\vert g\rangle \langle e\vert) +\Omega_{2}(
\vert e\rangle \langle f\vert +\vert f\rangle \langle e\vert),\nonumber\\   \label{HamitthreeVI}
\end{eqnarray}
where $\Delta_{1}=E_{e}-\omega_{1}$ and $\Delta_{2}=E_{e}-E_{f}-\omega_{2}$ are the transition detunings.

To better study the eigenvalues and eigenstates of the Hamiltonian~(\ref{HamitthreeVI}), we introduce three basis states defined by the vectors
\begin{equation}
\vert e\rangle =(
\begin{array}{ccc}
1 & 0 & 0
\end{array}
)^{T},\hspace{0.3 cm} \vert f\rangle =(
\begin{array}{ccc}
0 & 1 & 0
\end{array}
)^{T},\hspace{0.3 cm}\vert g\rangle =(
\begin{array}{ccc}
0 & 0 & 1
\end{array}
)^{T}.
\end{equation}
To demonstrate the dark-state effect in this three-level system, we consider the single- and two-photon resonance cases, i.e., $\Delta _{1}=\Delta _{2}=0$. Then the Hamiltonian $V_{I}$ can be written as
\begin{equation}
V_{I}=\Omega_{2}\left(
\begin{array}{ccc}
0 & 1  & \xi  \\
1  & 0 & 0 \\
\xi  & 0 & 0
\end{array}
\right), \label{matrixVI}
\end{equation}
where $\xi=\Omega_{1}/\Omega_{2}$ is the amplitude ratio.

The dark-state effect of this three-level system can be analyzed by calculating the eigensystem of the matrix $V_{I}$ given in Eq.~(\ref{matrixVI}). The eigen equation of $V_{I}$ reads
\begin{equation}
V_{I}\left\vert\lambda_{s}\right\rangle =\lambda_{s}\left\vert \lambda _{s}\right\rangle, \hspace{0.5 cm} s=1,2,3.
\end{equation}
The eigenvalues are given by $\lambda_{1}=0$, $\lambda_{2}=-\Omega_{2}\sqrt{1+\xi^{2}}$, and $\lambda_{3}=\Omega_{2}\sqrt{1+\xi^{2}}$.
The corresponding eigenstates can be obtained as
\begin{eqnarray}
\vert\lambda_{1}\rangle&=&\frac{1}{\sqrt{1+\xi^{2}}}(0\vert e\rangle-\xi\vert f\rangle+\vert g\rangle),\nonumber\\
\vert\lambda_{2}\rangle&=&\frac{1}{\sqrt{2(1+\xi^{2})}}\left(-\sqrt{1+\xi^{2}}\vert e\rangle+\vert f\rangle+\xi\vert g\rangle\right),\nonumber\\
\vert\lambda_{3}\rangle&=&\frac{1}{\sqrt{2(1+\xi^{2})}}\left(\sqrt{1+\xi^{2}}\vert e\rangle+\vert f\rangle+\xi\vert g\rangle\right).\label{eigenstates}
\end{eqnarray}
To study the dark-state effect in this system, we can calculate the probability of the excited state $|e\rangle$ in these eigenstates,
\begin{equation}
P_{e}^{[s]}=\vert\langle e\vert\lambda_{s}\rangle\vert^{2}.\label{eigenstatepe}
\end{equation}
By combining Eqs.~(\ref{eigenstates}) and (\ref{eigenstatepe}), we find that the probability of the eigenstate $\vert\lambda_{1}\rangle$ is always zero, which means that the eigenstate $\vert\lambda_{1}\rangle$ is the dark state.

To break the dark-state effect in the $\Lambda$-type three-level system, we introduce an auxiliary state coupled to the lower states $|f\rangle$, forming an $N$-type four-level system. The Hamiltonian of this system can be expressed as
\begin{eqnarray}
H_{\text{FLS}}(t)=H_{\text{TLS}}(t)+E_{d}|d\rangle\langle d|+\Omega_{3}(\vert d\rangle \langle f\vert e^{-i\omega_{3}t}+\mathrm{H.c.}),
\end{eqnarray}
where $\omega_{3}$ is the frequency of the monochromatic field and $\vert d\rangle$ is the auxiliary state coupled to the lower state $\vert f\rangle$ with the coupling strength $\Omega_{3}$. In a rotating frame with respect to $\omega_{1}\vert e\rangle \langle e\vert+E_{f}\vert f\rangle \langle f\vert +E_{g}\vert g\rangle \langle g\vert+(E_{f}+\omega_{3})\vert d\rangle\langle d\vert$, the Hamiltonian $H_{\text{FLS}}$ becomes
\begin{eqnarray}
V^{\prime}_{I}&=&\Delta_{1}\vert e\rangle \langle e\vert +\Delta_{3}\vert d\rangle \langle d\vert +\Omega_{1}(\vert e\rangle \langle g\vert +\vert
g\rangle \langle e\vert) \nonumber\\
&&+\Omega _{2}(\vert e\rangle \langle f\vert+\vert f\rangle \langle e\vert) +\Omega_{3}(
\vert d\rangle \langle f\vert +\vert f\rangle \langle d\vert),
\end{eqnarray}
where $\Delta_{3}=E_{d}-E_{f}-\omega_{3}$ is the transition detuning.

%%%%%%%%%%%%%%%%%%%%%
\begin{figure}[tbp]
\center
\includegraphics[width=0.47\textwidth]{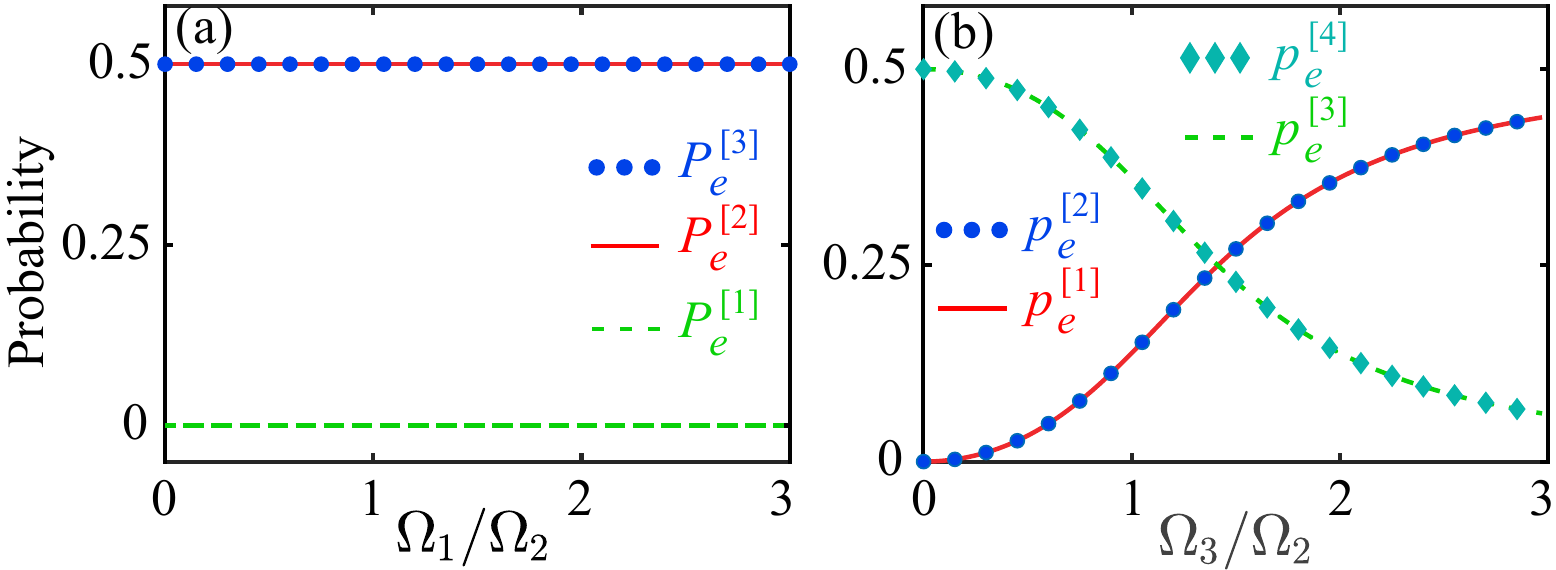}
\caption{(a) Probability $P_{e}^{[s]}$ of the excited state $|e\rangle$ in three eigenstates versus the amplitude ratio $\Omega_{1}/\Omega_{2}$ in the three-level system. (b) Probability $p_{e}^{[s]}$ of four eigenstates versus the amplitude ratio $\Omega_{3}/\Omega_{2}$ in the four-level system.}
\label{Fig13}
\end{figure}
%%%%%%%%%%%%%%%%%%%%%%%

By defining the four basis states with the vectors
\begin{eqnarray}
&&\vert e\rangle =(
\begin{array}{cccc}
1 & 0 & 0 & 0
\end{array}
)^{T},\hspace{0.5 cm}\vert f\rangle =(
\begin{array}{cccc}
0 & 1 & 0 & 0
\end{array}
)^{T},\nonumber\\
&&\vert g\rangle=(
\begin{array}{cccc}
0 & 0 & 1 & 0
\end{array}
)^{T},\hspace{0.5 cm}
\vert d\rangle=(
\begin{array}{cccc}
0 & 0 & 0 & 1
\end{array}
)^{T},
\end{eqnarray}
the Hamiltonian $V^{\prime}_{I}$ can be expressed as
\begin{equation}
V^{\prime}_{I} =\Omega^{\prime}\left(
\begin{array}{cccc}
0 & 1 & 1 & 0 \\
1 & 0 & 0 & \xi^{\prime} \\
1 & 0 & 0 & 0 \\
0 & \xi^{\prime} & 0 & 0
\end{array}
\right).
\end{equation}
Here, we introduce the parameter $\Omega_{1}=\Omega_{2}=\Omega^{\prime}$ and $\Omega_{3}/\Omega^{\prime}=\xi^{\prime}$, and consider the case $\Delta_{1}=\Delta_{3}=0$. Follow the standard procedure as performed in the three-level system, we can obtain the eigenvalues of the matrix $V^{\prime}_{I}$ as
\begin{equation}
\lambda_{s=1,2,3,4}^{\prime}=\pm\Omega^{\prime}\sqrt{\frac{2+\xi^{\prime2}\pm\sqrt{(2+\xi^{\prime2})^{2}-4\xi^{\prime2}}}{2}}.
\end{equation}
The corresponding eigenstates are given by
\begin{equation}
\vert\lambda_{s}^{\prime}\rangle =|\delta|[\lambda_{s}^{\prime}(\lambda_{s}^{\prime2}-\xi^{\prime2})\vert e\rangle+\lambda_{s}^{\prime2}\vert f\rangle+ (\lambda_{s}^{\prime2}-\xi^{\prime2})\vert g\rangle+ \lambda_{s}^{\prime}\xi^{\prime}\vert d\rangle],
\end{equation}
where $\vert\delta\vert^{2}=1/[(\lambda_{s}^{\prime2}+1)(\lambda_{s}^{\prime2}-\xi^{\prime2})^{2}+\lambda_{s}^{\prime2}(\lambda_{s}^{\prime2}+\xi^{\prime2})]$ is the normalization constant. To check the dark-state effect in this system, we also calculate the probability of these eigenstates,
\begin{equation}
p_{e}^{[s]}=\vert\langle e\vert\lambda_{s}^{\prime}\rangle\vert^{2},
\end{equation}
where $p_{e}$ is the probability of the excited state $|e\rangle$ in the $N$-type four-level system.

\begin{table*}[t]
\centering
\caption{Reported experimental parameters in the electromechanical system~\cite{Ockeloen-Korppi2018} and the scaled parameters used in our simulations. Columns 1, 2, 3, and 4 present the notation, the definitions, the reported experimental parameters, and the scaled parameters used, respectively.} \label{table1}
\begin{tabular*}{1\textwidth}{@{\extracolsep{\fill}}c c c c}
\toprule
Notation     & Definition    & Parameters in Ref.~\cite{Ockeloen-Korppi2018}   &  Parameters used  \\ \hline
$\omega_{1}$ &  frequency of the first mechanical oscillator       & $2\pi\times 10$ MHz   & frequency scale  \\ \hline
$\omega_{2}$ &  frequency of the second mechanical oscillator      & $2\pi\times 11.3$ MHz & $\omega_{2}/\omega_{1}=1$  \\ \hline
$\kappa$     & decay rate of the intermediate cavity field         & $2\pi\times1.38$ MHz  & $\kappa/\omega_{1}=0.1$   \\ \hline
$\kappa_{s}$ & decay rate of the  auxiliary cavity field           &                       & $\kappa_{s}/\omega_{1}=0.1$   \\  \hline
$G_{1 (2)}$   & effective optomechanical-coupling strength          & $2\pi\times (0.1\text{-}0.5)$MHz  & $G_{1 (2)}/\omega_{1}=0.05$  \\ \hline
$G_{s1 (s2)}$ & effective optomechanical-coupling strength          &                                   & $G_{s1 (s2)}/\omega_{1}=0.08$  \\ \hline
$n_{1}^{th}$ & phonon number in the first mechanical oscillator    & 41                     & 1000  \\ \hline
$n_{2}^{th}$ & phonon number in the second mechanical oscillator   & 30                     & 1000  \\  \hline
$\gamma_{1}$ & damping rate of the first mechanical oscillator     & $2\pi\times 106$ Hz               & $\gamma_{1}/\omega_{1}=10^{-5}$   \\  \hline
$\gamma_{2}$ & damping rate of the second mechanical oscillator    & $2\pi\times 144$ Hz               & $\gamma_{2}/\omega_{1}=10^{-5}$   \\  \hline
$\eta$       & phonon-hopping coupling strength                    &                                   & $\eta/\omega_{1}=0.03\text{-}0.06$ \\  \hline
$J$          & photon-hopping coupling strength                    &                                   & $J/\omega_{1}=0.03$ \\  \hline
\botrule
\end{tabular*}
\end{table*}

In Fig.~\ref{Fig13}(a) we plot the probability $P_{e}^{[s]}$ of these three eigenstates $|\lambda_{s}\rangle$ ($s=1$, $2$, $3$, and $4 $) as a function of the amplitude ratio $\Omega_{1}/\Omega_{2}$ in the three-level system. We can find that the probability $P_{e}^{[1]}$ is always zero no matter how the ratio changes, which means that the eigenstate $\vert\lambda_{1}\rangle$ is the dark state. In Fig.~\ref{Fig13}(b) we plot the probability $p_{e}^{[s]}$ of these four eigenstates $|\lambda_{s}^{'}\rangle$ as a function of the amplitude ratio $\Omega_{3}/\Omega_{2}$ in the $N$-type four-level system. Here we can see that the probabilities $p_{e}^{[1]}$ and $p_{e}^{[2]}$ are zero when the amplitude ratio $\Omega_{3}/\Omega_{2}=0$, which means that the two eigenstates $|\lambda_{1}^{\prime}\rangle$ and $|\lambda_{2}^{\prime}\rangle$ are dark states when the transition channel between the auxiliary state $|d\rangle$ and the lower state $|f\rangle$ is closed. However, with the increase of the ratio $\Omega_{3}/\Omega_{2}$, the excited-state probability of the four eigenstates is nonzero, which means that the dark-state effect is broken when the auxiliary state is introduced to the system.

\section{Discussion on experiment implementation\label{Discuss}}

In this section we present a discussion of the experimental implementation of this scheme. This system only involves the linearized optomechanical couplings and photon- or phonon-hopping coupling, which are experimentally accessible in current optomechanical systems~\cite{Aspelmeyer2014}. In the simulations, we consider the model in the resolved-sideband regime $\omega_{l=1,2}\gg\kappa_{(s)}$ and take the linearized coupling strengths as $G_{l}/\omega_{1}<0.1$ and $G_{sl}/\omega_{1}<0.1$ for $l=1,2$, which have been realized in many optomechanical systems~\cite{Aspelmeyer2014}. The photon-hopping and phonon-hopping interactions have been experimentally realized in optomechanical crystal circuits~\cite{Fang2017} and double-microdisk whispering-gallery resonators~\cite{Lin2010}. All these advances confirm the experimental feasibility of this scheme.

Next we present a parameter analysis based on the circuit electromechanical systems~\cite{Teufel2011,Ockeloen-Korppi2018,Ockeloen-Korppi2019,Lepinay2021,Kotler2021}, where the two effective microwave cavities are two superconducting circuits on a quartz substrate. They have the same resonance frequency $\omega_{c}= \omega_{s}\approx2\pi\times4.2$ GHz. The two mechanical modes are two drum resonators that function as compliant capacitances in the circuit. The mechanical position can affect the resonance frequency of microwave cavity; then the optomechanical interaction can be realized. Note that one of the two microwave cavities act as the intermediate coupling cavity mode coupled to two mechanical modes, while the other microwave cavity is only coupled to one of the two mechanical modes. Here the mechanical modes of two drums have the resonance frequency $\omega_{1}= \omega_{2}\approx2\pi\times 10$ MHz. The intrinsic energy decay rates of two microwave cavities and two drum resonators are $\kappa=\kappa_{s}\approx2\pi\times1$ MHz and $\gamma_{1}=\gamma_{2}\approx 2\pi\times100$ Hz, respectively. The effective coupling rates between the intermediate-coupling microwave cavity and the two drum resonators are $G_{1}=G_{2}\approx2\pi\times 0.5$ MHz and the effective coupling rates between the auxiliary microwave cavity and the two drum resonators are $G_{s1}=G_{s2}\approx 2\pi\times0.8$ MHz. In Table~\ref{table1} we present the reported experimental parameters in the circuit electromechanical systems~\cite{Ockeloen-Korppi2018} and the scaled parameters used in our simulations. By comparing these scaled parameters and the experimental parameters, we find that the experimental implementation of this scheme should be within the reach of current or near-future experimental conditions.

\section{Conclusion\label{Conclu}}

We have proposed an auxiliary-cavity-mode method to realize simultaneous ground-state cooling of two degenerate or nearly degenerate mechanical modes. We have also studied the general physical coupling configuration for breaking the dark mode in the network-coupled four-mode optomechanical system. The analytical parameter conditions for breaking the dark-mode effect have been found. Moreover, we have generalized this method to realize ground-state cooling of multiple mechanical modes, which was achieved by introducing the auxiliary cavity mode and phonon-hopping couplings between nearest-neighbor mechanical modes. We also described the physical mechanism for breaking the dark-state effect in the $N$-type four-level atomic system. Our results pave a way toward the demonstration of macroscopic quantum coherence and quantum manipulation in multiple-mechanical-mode optomechanical systems.

\begin{acknowledgments}
J.-Q.L. was supported in part by the National Natural Science Foundation of China (Grants No.~12175061, No.~11822501, No.~11774087, and No.~11935006), the Science and Technology Innovation Program of Hunan Province (Grants No.~2021RC4029 and No.~2020RC4047), and Hunan Science and Technology Plan Project (Grant No.~2017XK2018). J.-F.H. was supported in part by the National Natural Science Foundation of China (Grant No.~12075083) and Natural Science Foundation of Hunan Province, China (Grant No.~2020JJ5345). F.N. was supported in part by Nippon Telegraph and Telephone Corporation Research, the Japan Science and Technology Agency (via the Quantum Leap Flagship Program, Moonshot R$\&$D Grant No.~JPMJMS2061), the Japan Society for the Promotion of Science (JSPS) (via the Grants-in-Aid for Scientific Research Grant No.~JP20H00134), the Army Research Office (Grant No.~W911NF-18-1-0358), the Asian Office of Aerospace Research and Development (via Grant No.~FA2386-20-1-4069), and the Foundational Questions Institute Fund via Grant No.~FQXi-IAF19-06.
\end{acknowledgments}

\end{document}